 \documentclass[prb,a4paper,twocolumn,floatfix]{revtex4-1}
\usepackage{color}
\usepackage[usenames,dvipsnames]{xcolor}
\usepackage{graphicx}
\usepackage{latexsym}
\usepackage[]{amsmath}
\usepackage{epstopdf}
\usepackage{amssymb}
\usepackage{array}

\usepackage[utf8]{inputenc}
\usepackage[T1]{fontenc}
\usepackage{mathptmx}

\usepackage{dcolumn}
\usepackage{bm}

\begin{document}

\title[Information launched from accelerated polarization currents]{Information carried by electromagnetic radiation launched from
    accelerated polarization currents}

\author{John Singleton$^{\aleph}$}
\affiliation{National High Magnetic Field Laboratory, MPA-MAGLAB, MS-E536, 
Los Alamos National Laboratory, Los Alamos, NM~87545, U.S.A.}
\author{Andrea C. Schmidt$^{\aleph}$}
\affiliation{National High Magnetic Field Laboratory, MPA-MAGLAB, MS-E536, 
Los Alamos National Laboratory, Los Alamos, NM~87545, U.S.A.}
\author{Connor Bailey}
\affiliation{National High Magnetic Field Laboratory, MPA-MAGLAB, MS-E536, 
Los Alamos National Laboratory, Los Alamos, NM~87545, U.S.A.}
\author{James Wigger}
\affiliation{National High Magnetic Field Laboratory, MPA-MAGLAB, MS-E536, 
Los Alamos National Laboratory, Los Alamos, NM~87545, U.S.A.}
\author{Frank Krawczyk}
\affiliation{Accelerators and Electrodynamics, AOT-AE, MS-H851,
Los Alamos National Laboratory, Los Alamos, NM~87545,  U.S.A.}
\begin{abstract}
We show experimentally that a continuous, 
linear, dielectric antenna in which a superluminal 
polarization-current distribution
accelerates can be used to transmit a broadband signal that is reproduced 
in a comprehensible form at a chosen target distance and angle.
The requirement for this exact correspondence between broadcast and received signals
is that each moving point in the polarization-current distribution approaches the 
target at the speed of light at all times during its transit along the antenna.
This results in a one-to-one correspondence between the time at which each point on the moving 
polarization current enters the antenna and the time at which {\it all} of the radiation emitted by this particular point
during its transit through the antenna arrives simultaneously at the target. 
This has the effect of reproducing the desired time dependence of the
original broadcast signal.
For other observer/detector positions, the time dependence 
of the signal is scrambled, due to the non-trivial
relationship between emission (retarded) time and reception time.
This technique represents a contrast to conventional radio 
transmission methods;
in most examples of the latter, signals are broadcast with little or no directivity, selectivity 
of reception being achieved through the use of narrow frequency bands. 
In place of this, the current paper uses a spread of frequencies to transmit information 
to a particular location; the signal is weaker and has a scrambled time dependence elsewhere.
We point out the possible relevance of this mechanism to 5G neighbourhood networks
and pulsar astronomy.
\end{abstract}


\maketitle

\section{Introduction}
Though the subject has been studied for over a century~\cite{Sommerfeldt,Schott,Ginzburg},
in the past 20 years there has been renewed interest in the emission of radiation by 
polarization currents that travel faster than the speed of light {\it in 
vacuo}~\cite{bolo2006,bolo2005,Bess2004,Bess2006,ynL00,jap,IEEE}. 
Such polarization currents may be produced by
photoemission from a surface excited by an obliquely incident, 
high-power laser pulse~\cite{bolo2006,bolo2005,Bess2004,Bess2006,ynL00}.
Alternatively, in {\it polarization-current antennas}, they are excited by the application of carefully timed 
voltages to multiple electrodes on either side of a slab of a dielectric such 
as alumina~\cite{jap,IEEE,acS13,RadarPatent,FeedMk1,FeedMk2,CryptoPatent}.  
To illustrate these emission mechanisims, we write the third and fourth 
Maxwell Equations~\cite{Jackson,Balanis,Bleaney,Jefimenko} in the following form:
\begin{equation}
\nabla \times {\bf E} +\frac{\partial {\bf B}}{\partial t} = 0
\end{equation}
\begin{equation}
\nabla \times {\bf H}-\epsilon_0 \frac{\partial {\bf E}}{\partial t} = {\bf J}_{\rm free} +\frac{\partial {\bf P}}{\partial t}.
\label{M4}
\end{equation}
Here {\bf E} is the electric field, {\bf H} is the magnetic field, ${\bf B}~ [=\mu_0 ({\bf H} + {\bf M)}]$
is the magnetic flux density, {\bf M} is the magnetization,
{\bf P} is the polarization ({\it i.e.,} the dipole moment per unit volume) and ${\bf J}_{\rm free}$
is a current density of mobile charges.
The terms on the left-hand side of both expressions are coupled equations that describe
the propagation of electromagnetic waves~\cite{Jackson,Bleaney},
whereas the terms on the right-hand side of Eq. \ref{M4} may be 
regarded as {\it source terms}~\cite{Balanis,Jefimenko}.
The current density ${\bf J}_{\rm free}$ of free charges (usually electrons) is used to generate 
electromagnetic radiation 
in almost all conventional applications such as phased arrays 
and other antennas~\cite{Balanis}, synchrotrons~\cite{synchrotron},
light bulbs~\cite{Bleaney} {\it etc.}. By contrast, the emission mechanisms
mentioned above employ the polarization current density, $\frac{\partial {\bf P}}{\partial t}$,
as their source term~\cite{jap,IEEE,acS13,RadarPatent,FeedMk1,FeedMk2,CryptoPatent}.

In this paper, we use an experiment to study the information conveyed in the signals broadcast by such
polarization currents when they are accelerated. 
We find that a time-dependent amplitude modulation is reproduced exactly in the received
signal only when the detecting antenna is close to a particular set of points,
the position of which is related to details of the acceleration. At other 
points, the signal is scrambled.
The result has implications for communication applications 
and for astronomical observations of objects such as pulsars.
\begin{figure}[tbp]   
	\centering
	\includegraphics[height=7cm]{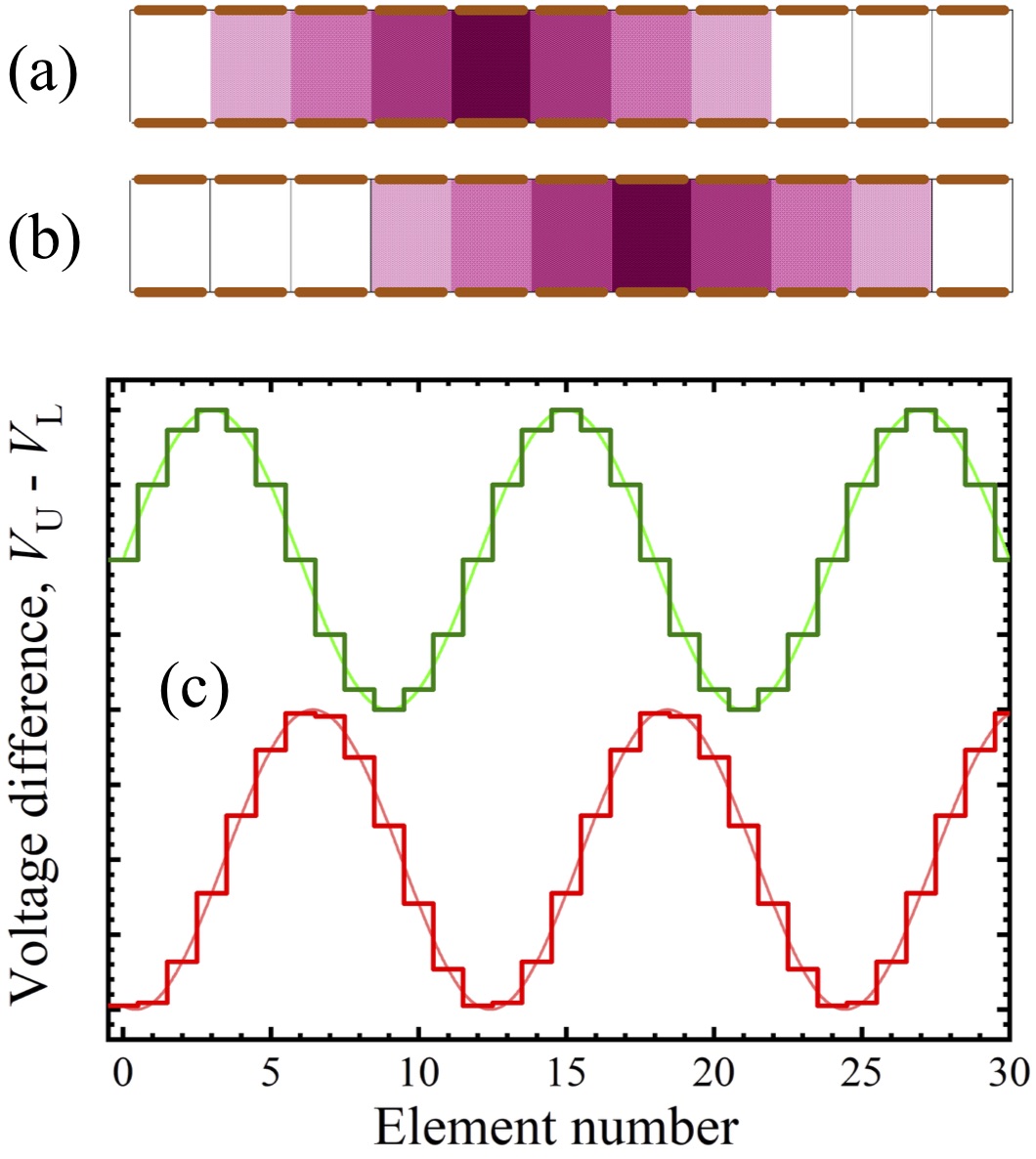}
	\caption{(a)~Polarization-current antennas (PCAs)
		consist of a continuous series of elements
		made of a dielectric(white)
		sandwiched between pairs of electrodes (orange).
		The dielectric is polarized by applying a voltage
		difference $V_{\rm U}-V_{\rm L}$ between
		upper (U) and lower (L) electrodes;
		this is shown schematically for seven
		elements, the shading density representing
		the polarization strength.
		(b)~The $V_{\rm U}-V_{\rm L}$
		shown in (a) are applied to elements two further
		to the right, moving the polarized
		region. 
		(c)~The upper (green) trace shows $[V_{\rm U}-V_{\rm L}]_j = 
		\sin[\omega(t-j\Delta t)]$ versus $j$, 
		where $j$ labels the antenna element, 
		$\omega$ is an angular frequency,
		$t$ is time and $\Delta t$ is a time increment, at $t=0$.
		The lower (red) trace (offset vertically for clarity)  
		shows $[V_{\rm U}-V_{\rm L}]_j$
		at $t=\frac{103}{180}\frac{2\pi}{\omega}$;
		the effect of the timing differences is to
		move the ``voltage wave'' (and the induced polarization) along.
		In practice, fringing effects round off the
		stepped voltages,
		leading to a smoother waveform (fine liines).}
	\label{figboom}
\end{figure}

The paper is organized as follows.
Section~\ref{animal} gives a brief introduction to the type of 
polarization-current antenna used in this work,
and how the polarization current within it is 
animated and accelerated; as the antennas may not
be familiar to the general reader, additional detail is given
in the Supplementary Information [SI]~\cite{SI}.
Section~\ref{accel} gives an account of the
acceleration scheme for transmitting information to particular locations.
Sections~\ref{RishiExpt} and \ref{RishiResults} describe an experimental proof-of-concept of
the effect carried out within a 6.5 m RF anechoic chamber.
Finally, Section~\ref{Rishi} discusses the implications of this 
observation for communications and astronomy.

\section{Polarization-current antennas}
\label{animal}
In both dielectric-resonator antennas (DRAs)~\cite{dielectric} 
and polarization-current antennas (PCAs)
dielectrics play a major role in the emission mechanisms.
However, the two antenna types function in completely different ways;
DRAs essentially use the dielectric
to boost the effective size (and hence the efficiency) of a small
antenna~\cite{dielectric}, whereas in PCAs, the dielectric hosts
a moving, volume-distributed polarization current~\cite{jap,IEEE,acS13,RadarPatent,FeedMk1,FeedMk2,CryptoPatent}.
Consequently, PCAs usually consist of a 
continuous strip of a dielectric
such as alumina with electrodes on
either side [Fig.~\ref{figboom}(a)]. 
Each electrode pair and the 
dielectric in between is referred to as an {\it element};
the elements are supplied independently
with a voltage difference, $V= V_{\rm U}-V_{\rm L}$,
where U and L refer to upper and lower electrodes.
This produces polarization {\bf P} in the  dielectric.
By changing $V_{\rm U}-V_{\rm L}$
on a series of elements, the polarized region is
moved [Fig.~\ref{figboom}(a),~(b)];
owing to the time dependence imparted by movement,
a polarization current, $\partial {\bf P}/\partial t$
is produced, and
will, under the correct conditions,
emit electromagnetic radiation~\cite{jap,IEEE,acS13,RadarPatent,FeedMk1,FeedMk2,CryptoPatent,Ginzburg,Schott}.

PCAs are usually run
by moving a continuous polarization current
along the dielectric~\cite{RadarPatent,FeedMk1,FeedMk2,CryptoPatent}..
This is accomplished by applying phase-shifted 
time-dependent signals
to the elements~\cite{SI}.
A simple example is given in Fig.~\ref{figboom}(c), 
where the upper (green) trace 
shows $[V_{\rm U}-V_{\rm L}]_j = 
\sin[\omega(t-j\Delta t)]$ versus $j$, 
where $j$ labels the antenna element, $\omega$ is an angular frequency,
$t$ is time and $\Delta t$ is a time increment, at $t=0$.
The lower (red) trace 
shows $[V_{\rm U}-V_{\rm L}]_j$
at a later time;
the effect of the time increments is to
move the ``voltage wave'' and hence the induced polarization
at a speed $v=a/\Delta t$,
where $a$ is the distance between element centres.
Acceleration is introduced by varying
$\Delta t$ along the antenna's length.
Further details and typical emission properties
are given in the SI~\cite{SI}.

\begin{figure}[htbp]
	\centering
	\includegraphics [width=0.85\columnwidth]{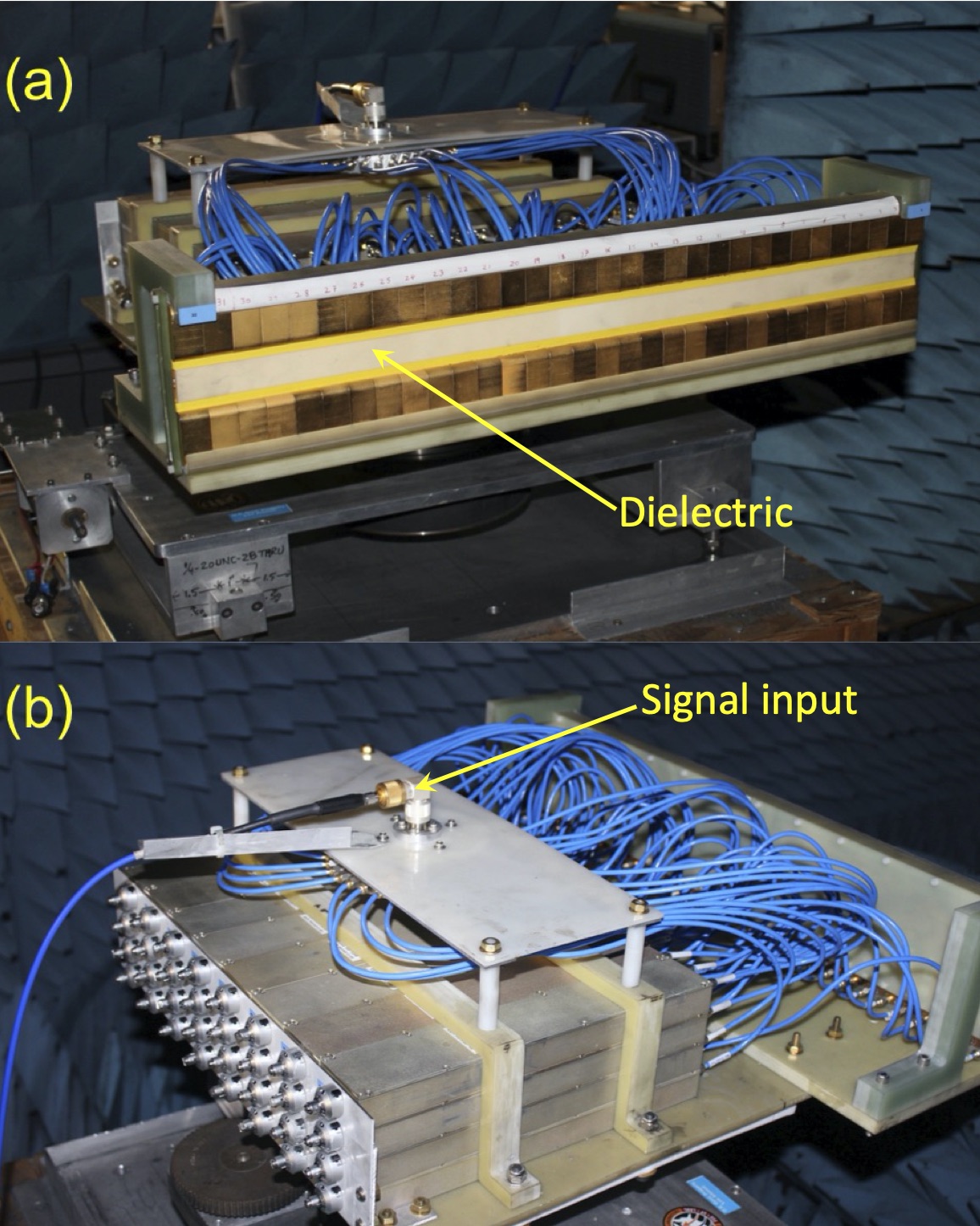}
	\caption{(a)~Front view of the passive antenna used in the
		demonstration experiment mounted on its turntable 
		in the anechoic chamber. It has 32 elements
		spanning a total length of 0.64~m.
		The label indicates the cream-colored dielectric (alumina) that hosts the
		volume-distributed moving polarization current responsible for
		the emission of radio waves from the antenna.
		(b)~Rear view of the antenna showing the 32-way 
		splitter feeding 32 independent ATM P1214 mechanical phase shifters.
		The dials for adjusting the phase are visible on the lower left of the picture.
		The signal input on the top of the 32-way splitter is labelled.}
	\label{ExpAntenna}
\end{figure}
The practical antenna used in the experiments below
is shown in Fig.~\ref{ExpAntenna}(a);
it has 32 elements spanning a total length of 0.64~m,
and the dielectric is alumina $(\epsilon_{\rm r} \approx 10)$.
The elements are fed via a 32-way splitter and
32 mechanical delay lines [Fig.~\ref{ExpAntenna}(b)]
which are adjusted to produce time differences $\Delta t$~[\onlinecite{FeedMk2}].
Note that in these antennas, the polarization current fills the entire dielectric; 
it is a continuously moving volume source of radiation that emits from an 
extended volume, rather than at a series of points or lines (as in a phased array).
Despite the discrete nature of the electrodes, simulations of 
our antennas performed with off-the-shelf electromagnetic 
software packages such as {\it Microwave Studio} show that fringing fields of adjacent 
electrode pairs lead to a voltage phase that 
varies slightly under the electrode~\cite{frank}; 
{\it i.e.,} the phase is more smoothly varying along the length of the antenna than the 
discrete arrangement of electrodes suggests~\cite{SI,dissert}.
This is represented by the smoother curves in Fig.~\ref{figboom}(c).
\section{Concept: acceleration and focusing}
\label{accel}
\begin{figure*}[htbp]
	\centering
	\includegraphics [width=1.5\columnwidth]{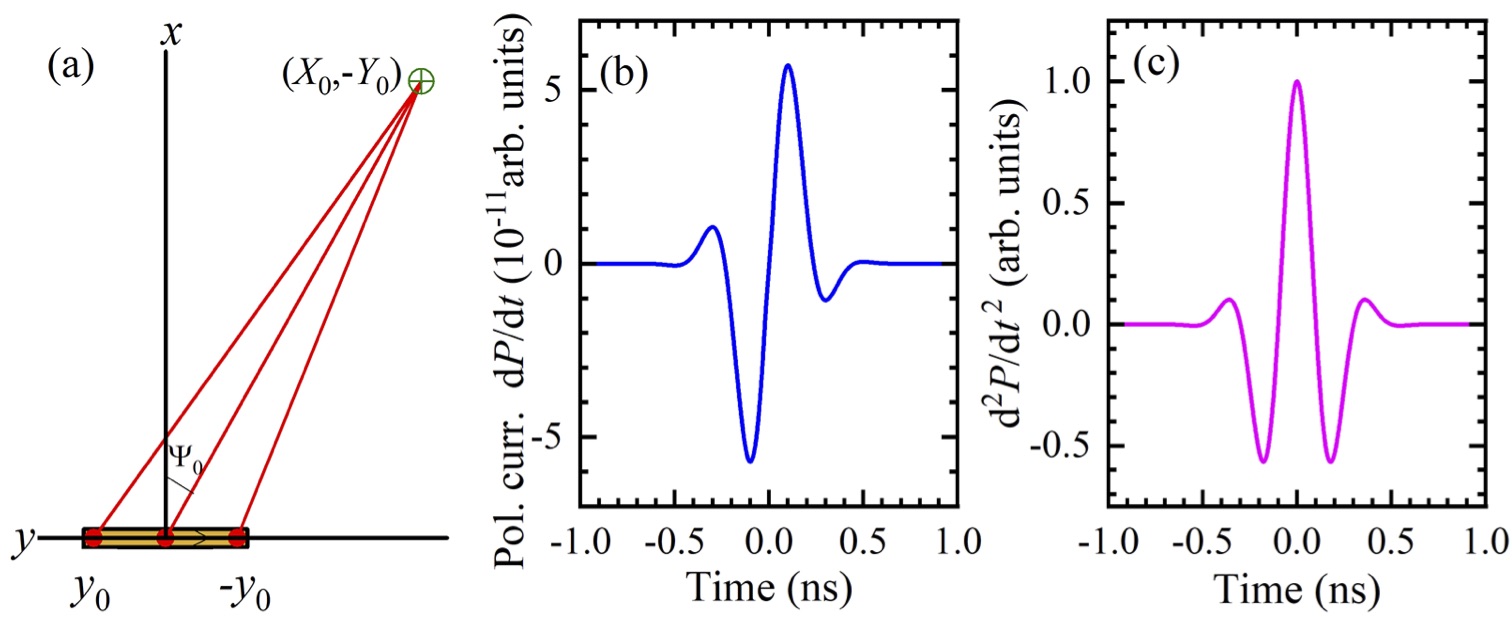}
	\caption{(a)~Experimental concept. An element of
		polarization current (red) moves along the dielectric
		antenna (dark yellow shading) such that the component
		of its velocity in the direction of the target [green cross
		at $(X_0, -Y_0)$] is always $c$,
		the speed of light. The center of the antenna is
		at  $(0,0)$.
		(b)~Notional time dependence of the polarization
		current ${\rm d}P/{\rm d}t$ sent along the antenna.
		(c)~Derivative of the curve shown in (b) with respect
		to time $t$. The ``arbitrary units'' in (c) are equivalent to those
		in (b), the large scaling occurring because of the fast time
		dependences.
	}
	\label{Concept}
\end{figure*}
We now consider an antenna containing a ``wavepacket'' of polarization current
that has finite extent in both space and time; it moves on a linear trajectory and accelerates.
Fig.~\ref{Concept}(a) shows a plan view of the 
antenna's dielectric
of length $2y_0$ with its center at $(0,0,0)$
lying along the Cartesian $y-$axis.
As in the experimental antennas~\cite{FeedMk1,FeedMk2,CryptoPatent} 
[Fig.~\ref{ExpAntenna}(a)]
the dielectric has rectangular cross-section;
its depth $2x_0$ (extent in the $x$ direction)
and height $2z_0$ (extent in the $z$ direction) are symmetrical
about the $y$ axis; both $x_0$ and $z_0$ are $\ll y_0$.

A target  
is chosen in the $(x,y)$ plane at a distance $R_0$;
the angle $\Psi_0$ ``off boresight'' 
describes the target's azimuthal position.
As everything of interest lies in the $(x,y)$ ($z=0$)
plane, for convenience we drop the Cartesian $z$
coordinate for the time being.
Thus, the target is at $(X_0,-Y_0)$, where 
\begin{equation}
X_0 =R_0 \cos \Psi_0 ~~~~{\rm and}~~~~|Y_0| = R_0 \sin \Psi_0.
\label{dawg}
\end{equation}
Consider a point in the polarization current
that is moving through the dielectic
along the $y-$axis;
the instantaneous distance $r$ between the point at $(0,y)$ 
and the target at $(X_0,Y_0)$ is given by
\begin{equation}
r^2 = X_0^2 + (Y_0+y)^2.
\label{reqn}
\end{equation} 
The point is made to move in such a way that
the component of its velocity towards the target
is always $c$, the speed of light in the surrounding medium
(assumed to be vacuum),
that is $({\rm d} r/{\rm d}t) = -c$, where $t$ is the time.
Differentiating Eq.~\ref{reqn} with respect to $t$,
inserting the above value for $({\rm d} r/{\rm d}t)$ and rearranging,
we obtain the point's velocity along $y$:
\begin{equation}
\frac{{\rm d} y}{{\rm d} t} = -c\frac{\left[X_0^2 + (Y_0+y)^2\right]^{\frac{1}{2}}}{Y_0+y}.
\label{gonzales}
\end{equation}
Integrating Eq.~\ref{gonzales}, and assuming that
the point commences its journey along the antenna 
at $y=y_0$ and time $t=0$,
we obtain a relationship between the point's
position $y$ and time $t$:
\begin{equation}
t=\frac{1}{c}\left[(X_0^2+(Y_0+y_0)^2)^{\frac{1}{2}} -
(X_0^2+(Y_0+y)^2)^{\frac{1}{2}}\right]. 
\label{Dawg2}
\end{equation}
We now consider a detector placed at
a general point P with coordinates $(X,-Y)$ in the $(x,y)$ plane.
The radiation emitted by
the point as it travels along the antenna
will reach P at a time $t_{\rm P}$
given by
\begin{widetext}
\begin{equation}
t_{\rm p} = t + \frac{1}{c}(X^2+(Y+y)^2)^{\frac{1}{2}} =\frac{1}{c}\left[(X_0^2+(Y_0+y_0)^2)^{\frac{1}{2}} -
(X_0^2+(Y_0+y)^2)^{\frac{1}{2}} 
+ (X^2+(Y+y)^2)^{\frac{1}{2}}\right]. 
\label{Biden}
\end{equation}
\end{widetext}
It should be obvious that if, {\it and only if}, $X=X_0$
and $Y=Y_0$, then $t_{\rm P} =$ a constant.
For all other choices of detector position,
$t_{\rm P}$  is a function of $y$ and therefore of $t$.
\begin{figure}[htbp]
	\centering
	\includegraphics [width=0.99\columnwidth]{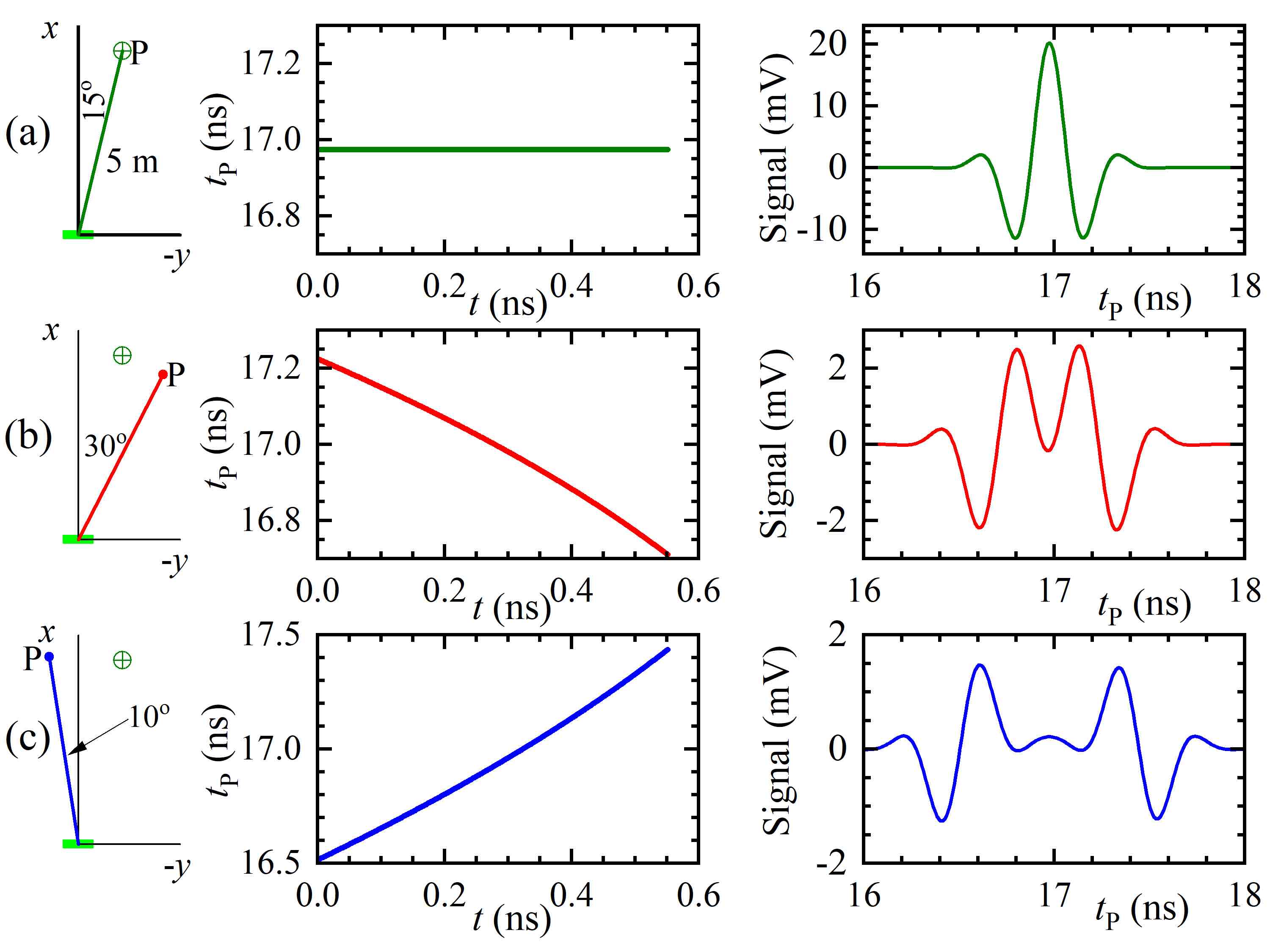}
	\caption{Each row shows the effect of
		moving the detector to positions P
		5~m from the antenna center (light green).
		In each row, the left panel gives position P,
		the center shows the 
		arrival time $t_{\rm P}$
		of radiation emitted at time $t$ by a point accelerating along 
		the antenna, and the right is the signal detected
		due to the polarization current of Fig.~\ref{Concept}(b)
		being accelerated along the antenna.
		In all cases, the target (green cross) is 5~m
		from the antenna center at an angle of $15^{\circ}$
		to the $x$ axis.
		Row~(a): detector at target position.
		Row~(b): detector placed on a line making
		an angle of $30^{\circ}$ with the $x$ axis.
		Row~(c): detector placed $10^{\circ}$ to the other side of the
		$x$ axis.}
	\label{ChirpFigure}
\end{figure}
\begin{figure*}[htbp]
	\centering
	\includegraphics [width=1.3\columnwidth]{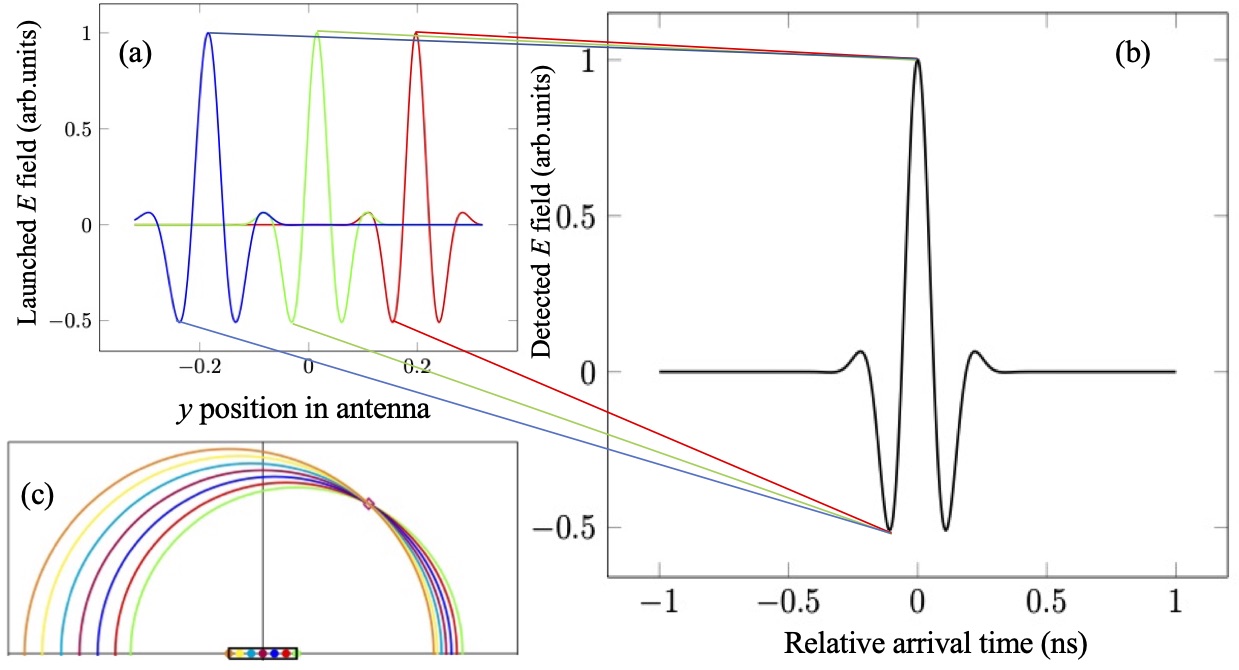}
	\caption{(a)~Launched $E-$field [$\propto ({\rm d}^2P/{\rm d}t^2)$],
	corresponding to a polarization current similar to that in Fig.~\ref{Concept}(b),
	at three different times (denoted by red, green and blue curves) 
	during its transit along the antenna; note that the waveform
	``stretches out’’ due to the acceleration parameterized by Eqs.~\ref{dawg} and \ref{Dawg2}.
	(b)~The corresponding detected $E-$field at the target point;
	colored lines linking (a) and (b) show schematically the principle that radiation from a particular point 
	on the traveling waveform arrives at the same time at the target. This is because the
	acceleration compensates exactly for the different distances between source point and 
	detection locations (see Eq.~\ref{Biden} {\it et seq.}).
	Hence, features in the launched $E$-field are reinforced in the correct time sequence 
	in the detected signal.
	(c)~The same principle is illustrated using Huyghens wavelets;
	colored dots in the antenna (dielectric outlined by black lines) 
	represent the positions of a particular point on waveform~(a) at different times 
	during its transit of the antenna; semicircles of the same color show corresponding emittted
	Huyghens wavelets arriving at the focus point (orange diamond) simultaneously.}
	\label{Explain}
\end{figure*}

This situation is illustrated in the first two columns
of Fig.~\ref{ChirpFigure}.
The intended target $(X_0,-Y_0)$ is at $R_0= 5$~m from the antenna center and at $\Phi_0 =15^{\circ}$
(Fig.~\ref{ChirpFigure}, row (a), left column);
if the detector P is placed {\it exactly} at this position,
then $t_{\rm P}= {\rm constant}$ (row (a), center column).
The constant here is the transit time of light from
$y=y_0$, the place at which the point source enters the antenna
at $t=0$, to the target; 
subsequently the accelerated motion of the point
source along the antenna
exactly compensates for the changing point-to-target distance.
If, on the other hand, the detector position P is not
at $(X_0,-Y_0)$ [Fig.~\ref{ChirpFigure}, rows (b) and (c)],
then $t_{\rm P}$ is a function of $t$.

Next, rather than a single point, we consider the movement of the whole time-dependent 
polarization-current waveform along the antenna~\cite{footnotexz}. 
The imposed motion is such that {\it each} point within the waveform is accelerated as described above; 
{\it i.e.} as it traverses the antenna, such a point always has a velocity component $c$ in 
the direction of the target. Referring to the discussion of Eq.~\ref{Biden} above, all radiation emitted by this point as 
it moves along the antenna will arrive at the target at a time given by 
$t_{\rm P} =($time that point enters the antenna at $y=y_0)+($transit time of light from $y=y_0$ to the target). 
Therefore, there is a one-to-one correspondence between the time at which each point on the moving 
waveform enters the antenna and the arrival time at the target of the radiation emitted by this particular point
as it traverses the antenna. 

To show how this affects the received radiation, 
we send the waveform shown in Fig.~\ref{Concept}(b) 
along the antenna with the constraint that each point on the waveform obeys 
the acceleration scheme described by 
Eqs.~\ref{dawg} and \ref{Dawg2}; as before $\Phi_0 = 15^{\circ}$ and $R_0 =5.0$~m. 
The resulting signals (proportional to the $E-$field)
for the detector positions given in the first column of Fig.~\ref{ChirpFigure} 
are shown in the third column of
the same figure; the SI describes how such calculations are carried out~\cite{SI,dissert}. 
At the target angle and distance [(Fig.~\ref{ChirpFigure}, row (a)], the detected signal
reproduces the shape of the time derivative of the 
polarization-current waveform [Fig. 3(c)] exactly. Away from the target position [Fig. 4, rows (b) and (c)], 
the detected signal is much smaller and has altered frequency content and shape.

First, why is the time-derivative of the polarization current reproduced? The calculations in the SI show that~\cite{SI,dissert} 
the magnetic vector potential {\bf A} resulting from each volume element of the antenna is proportional 
to the polarization current within that element (SI, Eq.~8~[\onlinecite{SI}]). The corresponding 
$E-$field is proportional to the derivative of ${\bf A} $ with respect to time~\cite{Bleaney}. 
Therefore, it is the {\it electric field} launched from the antenna that is reproduced at, 
and only at,  the target point.

The idea is illustrated in more detail in Fig.~\ref{Explain}; (a) shows 
the launched $E-$field [$\propto ({\rm d}^2P/{\rm d}t^2)$]
at three different times indicated by different colors
during its transit through the antenna. 
Figure~\ref{Explain}(b) shows the corresponding detected $E-$field at the target point. 
The colored lines linking the curves in Figs.~\ref{Explain}(a) and (b) illustrate the principle that radiation 
from a particular point on the traveling waveform {\it always arrives at the same time at the target}. 
Thus, features in the launched $E$-field are {\it reinforced} at the target in the correct time sequence.
In other words, the time dependence of the emission of the whole waveform
is reproduced at the target [compare Figs~\ref{Concept}(c) and~\ref{ChirpFigure}(a)]
whereas elsewhere, it is scrambled [Figs.~\ref{ChirpFigure}(b), (c)]. 

Fig.~\ref{Explain}(c) illustrates the same principle using Huyghens wavelets. 
The colored dots represent the positions of a particular point on the polarization-current waveform 
at different times during its transit of the antenna; semicircles of the same color 
represent the corresponding emitted Huyghens wavelets, arriving at the target (orange diamond) 
simultaneously. At other locations, the Huyghens wavelets arrive at different times, 
so that the signal becomes scrambled.

For the experimental demonstration below,
we need to describe a polarization-current waveform the possesses the required 
motion for the above focusing effects.
To do this, we write~\cite{footnotexz}
\begin{equation}
\frac{\partial{\bf P}}{\partial t} = {\bf f}(y,t) = {\bf f}[t-p(y)],
\label{neepneep}
\end{equation}
where {\bf f} is a vector function of time $t$ and the function $p(y)$.
Constant phase points are represented by
$t-p(y) = {\rm constant}$; differentiating this with respect to $t$ results in
\begin{equation}
1 = \frac{{\rm d}p}{{\rm d}t} = \frac{{\rm d}p}{{\rm d}y}\frac{{\rm d}y}{{\rm d}t}
\end{equation}
Substituting from Eq.~\ref{gonzales} and integrating,
we obtain
\begin{equation}
p(y) = - \frac{\left[X_0^2+(Y_0+y)^2\right]^{\frac{1}{2}}}{c}.
\label{wileyecoyote}
\end{equation}
Eqs.~\ref{neepneep} and~\ref{wileyecoyote} describe
the required extended polarization current waveform, 
all of the points within which
approach the target at a speed of $c$.

Finally, note that we have only treated a time-domain focus in the $(x,y)$ plane.
In fact the criterion for focusing - that points in the 
polarization-current distribution approach the observer/detector
at the speed of light along their entire path though the antenna -
is fulfilled on a {\it semicircle} of points around the antenna [our antennas are
designed not to emit from their rear surfaces~\cite{FeedMk2}] that extends in the
$y$ and $z$ directions,
with a radius $(y^2+z^2)^{1/2} =Y_0$.
However, in a proof-of-concept
demonstration experiment, moving the observer/detector
away from $z=0$
complicates matters, as the radiation's $E$-field is no longer vertically polarized;
there is an additional component polarized parallel to $y$ [this may be deduced from the
calculations in the SI~\cite{SI}; for more details see Chapter 6 of
Ref.~\onlinecite{dissert}]. 
In the next implementation of this concept, the single antenna
discussed in the present paper is replaced
by an array of linear antennas configured to allow full three-dimensional $(x,y,z)$
control of the information focus point, along with minimization of the
parasitic $y$ polarization of the $E$-field~\cite{nichols}.
\section{Experimental demonstration}
\label{RishiExpt}
The antenna shown in Fig.~\ref{ExpAntenna} is used for the experimental demonstration.
It is mounted on a powered turntable (vertical rotation axis) with an azimuthal angular
precision of $\pm 0.1^{\circ}$. A Schwarzbeck-Mess
calibrated dipole at the same vertical height is used to receive the vertically polarized
transmitted radiation;
this is mounted on a TDK plastic tripod 
on rails that allows it to be moved to different 
distances without changing the height or angular alignment of the
equipment. The entire system is in a $5.8\times 3.6\times 3.6$~m$^3$ 
metal anechoic chamber completely lined with ETS-Lindgren EHP-12PCL 
pyramidal absorber tiles.

Signals received by the dipole are sent either
to a Hewlett-Packard HP8595E spectrum analyzer to monitor
power at a chosen frequency,
or to a Mini-Circuits TVA-82-213A broadband amplifier that
allows the time-dependent voltage to be viewed and/or
digitized using a Tektronix TDS7404 digital oscilloscope. 
Care is taken to ensure that the cables used
are shielded from the radiation
within the anechoic chamber and that secondary-path signals
are $\sim 60$~dB less than direct
radiation from antenna to dipole.

The description in Section~\ref{accel} is framed in terms of a traveling wavepacket. 
However, detecting a single pulse, especially if it contains a spread of frequencies, 
presents technical difficulties in a facility where only low power levels are permitted. 
Instead, we choose to transmit and detect what is in effect a train of wavepackets.
This forms a continuous broadband signal with a distinctive shape, based on a mixture 
of harmonics of 0.90~GHz and synthesized by mixing outputs from phase-locked 
TTi TGR6000 and Agilent N9318 function generators. 
\begin{figure}[htbp]
	\centering
	\includegraphics [width=0.8\columnwidth]{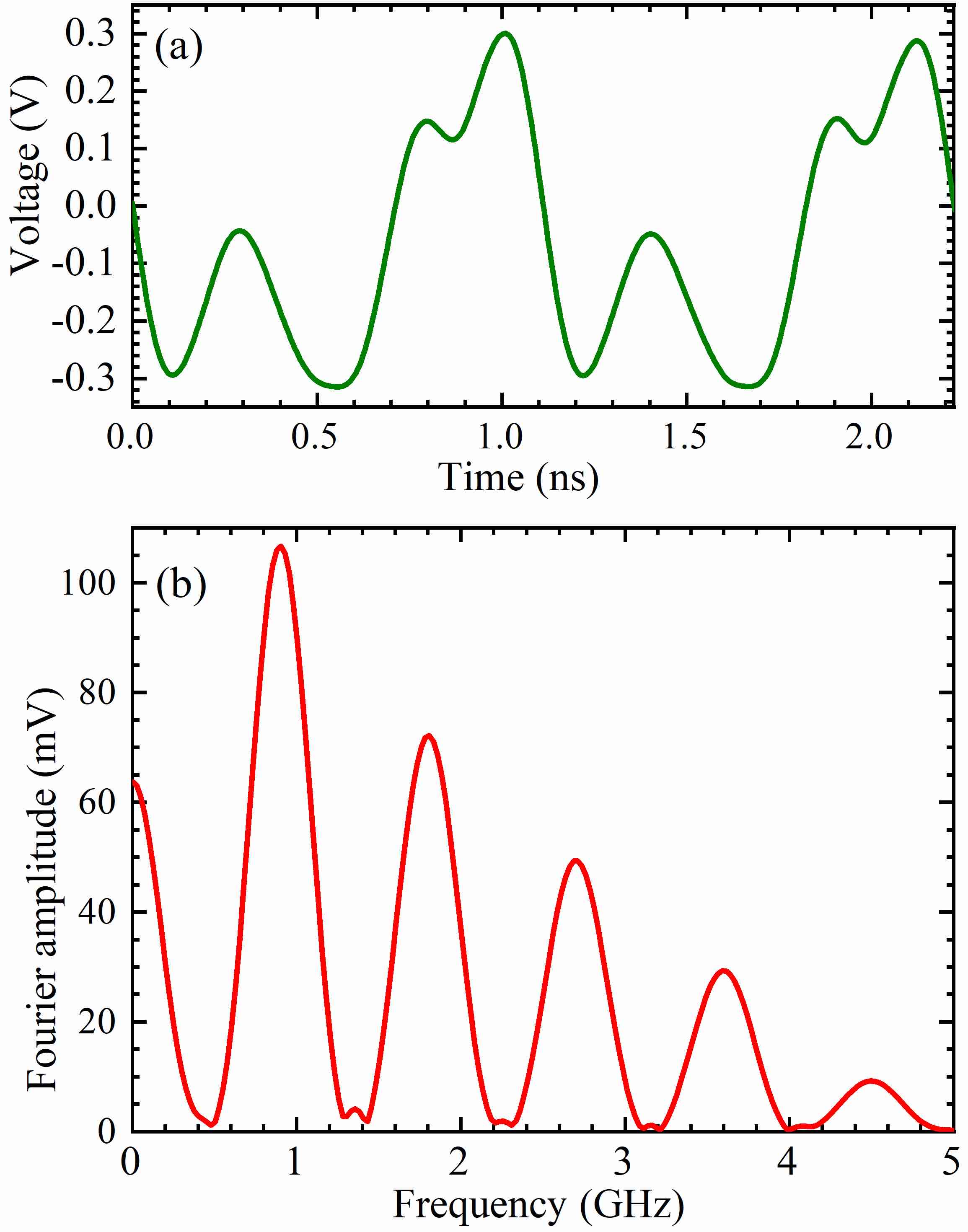}
	\caption{(a)~The voltage measured 
		by placing the dipole receiver 10~mm in front of the 16th 
		antenna element as a function of time;
		in effect, this is the desired transmitted signal.
		(b)~Fourier transform of the waveform in~(a).
		Note the distinctive ``triangular'' pattern of 
		harmonics of 0.9~GHz.}
	\label{waveform}
\end{figure}
The synthesized signal is sent to a Mini-Circuits TVA-82-213A amplifier, the 
output of which drives a 32-way splitter feeding 32 independent ATM P1214 
mechanical phase shifters [Fig.~\ref{ExpAntenna}(b)]. 
The latter are used to set the time-delays of the signals sent to each antenna element, 
reproducing the above acceleration scheme. 
To keep the ``information focus’’ well within the anechoic chamber, 
$X_0 =3.03$~m and $Y_0 = 0.64$~m are chosen, yielding target distance 
$R_0=3.09$~m and azimuthal angle $\Phi_0 = 11.9^{\circ}$. 

\begin{figure}[htbp]
	\centering
	\includegraphics [width=0.85\columnwidth]{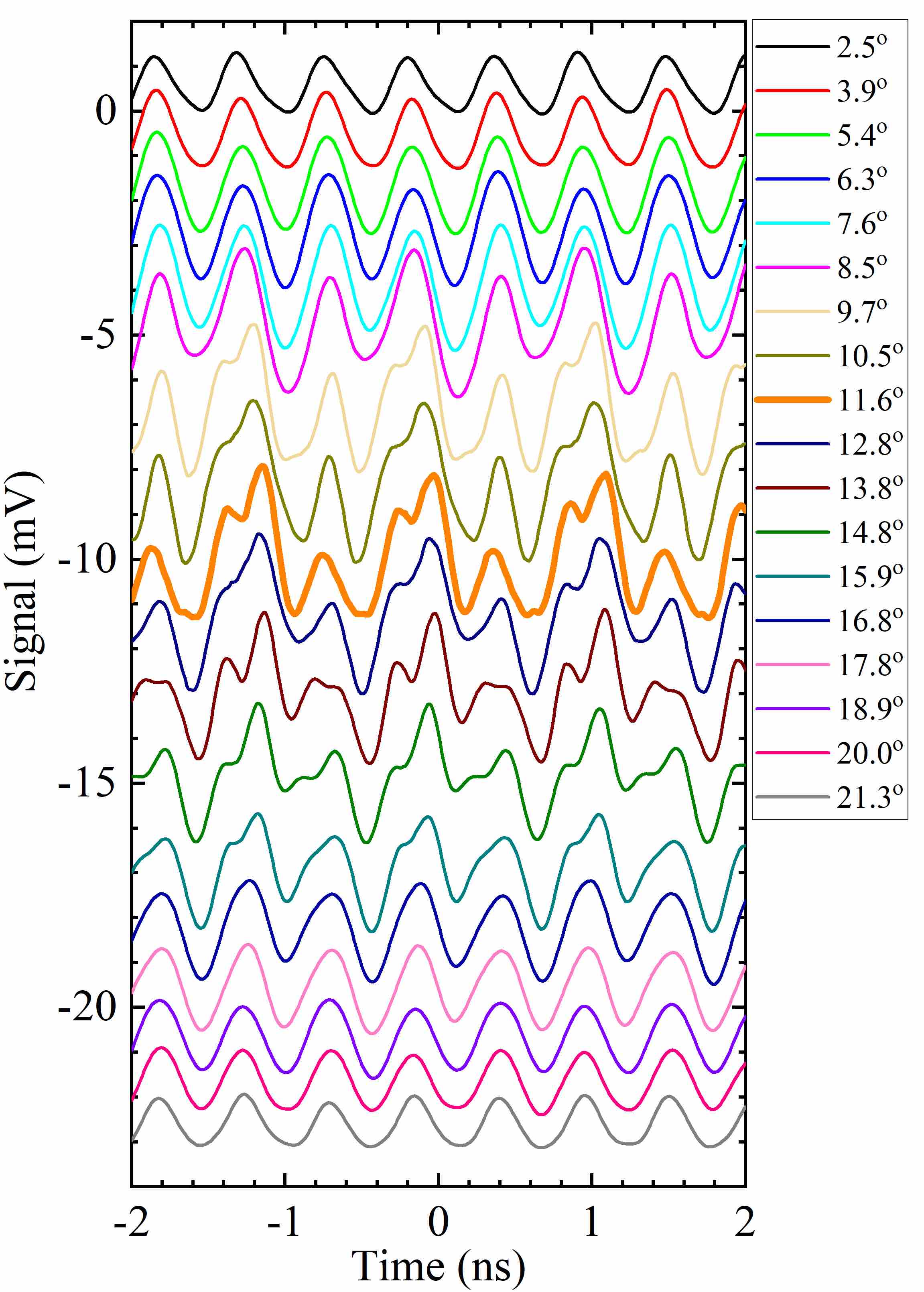}
	\caption{Time dependence of the signal received by the
		detector dipole for an antenna-to-detector distance
		of 3.0~m and for the azimuthal angles shown in the
		key. The shape of the transmitted waveform 
		(Figure~\ref{waveform}(b)) is only reproduced close to 
		the ``target'' angle of $11.6^{\circ}$ (orange trace).
		Experimental traces are offset vertically for clarity.}
	\label{timedepvsangle}
\end{figure}

The time-dependence of the broadcast waveform is recorded by placing the receiver 
dipole 10~mm in front of the 16th element of the antenna
and observing the signal on the oscilloscope. 
As long as the shortest emitted wavelength is much larger than distance
from the dielectric to the detector, the calculations described in the SI~\cite{SI} can 
be used to show that
the $E$-field thus detected by the dipole
is, to a good approximation,
$\propto \partial^2 {\bf P}/\partial t^2$, where ${\bf P}$ 
is the polarization passing the point
in the dielectric closest to the detector antenna.
Hence, an analogue of Fig.~\ref{Concept}(c) for the experimental 
wavetrain is captured; moreover, any frequency-dependent artefacts 
are the same in the measurements of the broadcast and received signals, 
making a comparison analogous to that between
Fig.~\ref{Concept}(b) and the third column of Fig.~\ref{ChirpFigure} simpler.
 
The waveform used for the experiments is selected by adjusting the 
outputs of the two signal generators and is shown in Fig.~\ref{waveform}(a). 
It is chosen because (i)~it has a distinctive time-dependent shape ({\it e.g.} the 
double peak followed by two differing minima, one relatively broad) and 
(ii)~an easily recognized ``triangular’’ Fourier spectrum [Fig.~\ref{waveform}(b)]. 
These traits aid in the rapid location of ranges of distance and azimuthal 
angle over which the broadcast signal was reproduced.
\begin{figure}[htbp]
	\centering
	\includegraphics [width=0.9\columnwidth]{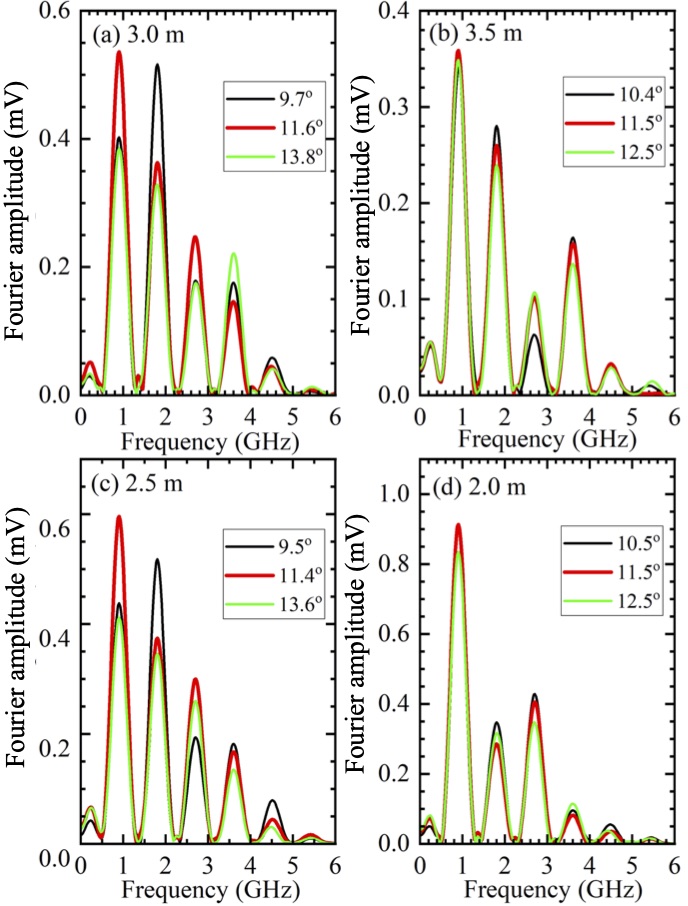}
	\caption{Fourier transforms of the detector signal for azimuthal angles (shown in key) 
	on either side of (green and black) and close to or at the target angle of $11.6^{\circ}$ (red) and at different antenna to detector distances:
	(a)~3.0~m, (b)~3.5~m, (c)~2.5~m and (d)~2.0~m.}
	\label{FTScramble}
\end{figure}

\section{Results}
\label{RishiResults}
\begin{figure*}[tbp]
	\centering
	\includegraphics[height=10cm]{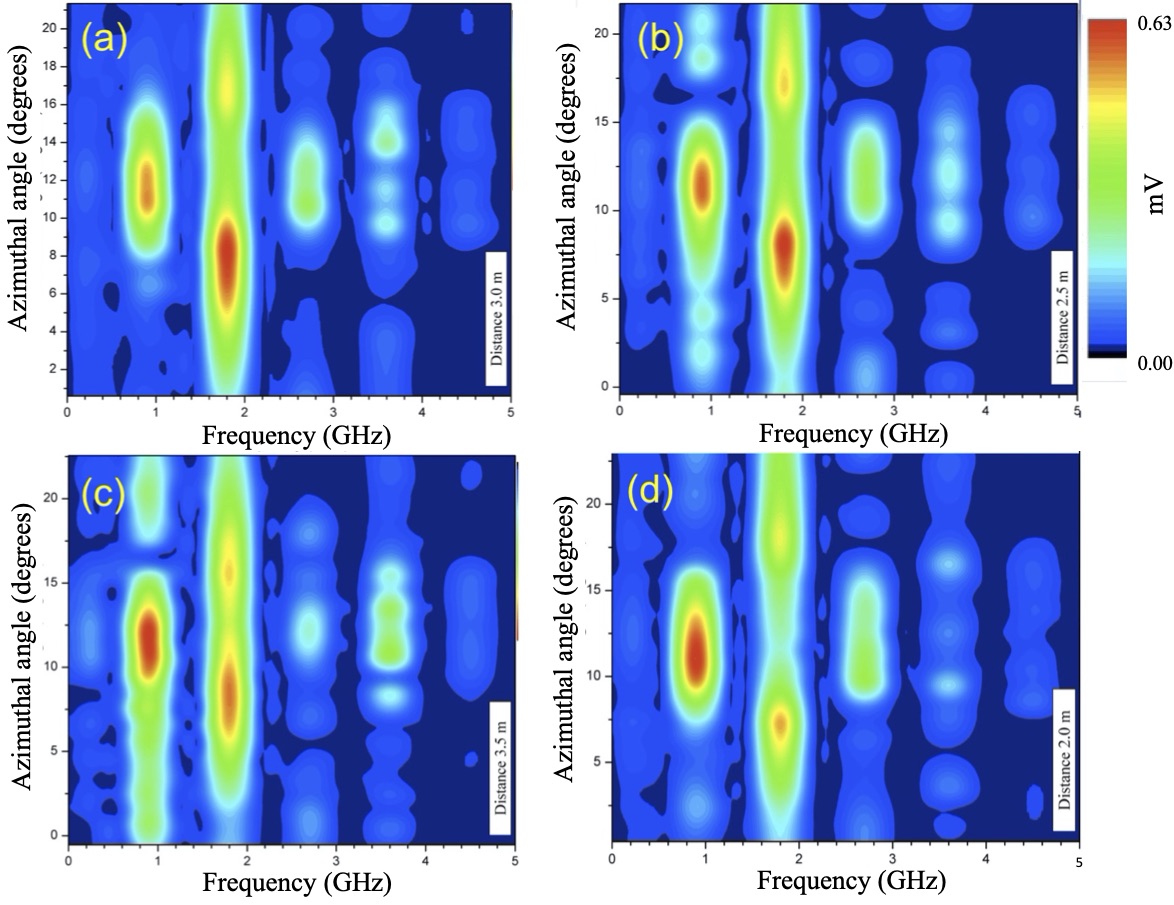}
	\caption{Fourier transforms of the detector signal 
		plotted as contour plots versus frequency and azimuthal angle for different antenna to detector distances:
		(a)~3.0~m, (b)~2.5~m, (c)~3.5~m and (d)~2.0~m.}
	\label{FTContour}
\end{figure*}

Preliminary surveys are carried out by sweeping the transmitter azimuthal 
angle at closely spaced distances around the expected $R_0$ whilst carefully 
observing the received signal on the oscilloscope or spectrum analyzer. 
Slight phase-setting errors result in actual target coordinates 
$R_0\approx 3.00$~m and $\Phi_0 \approx 11.6^{\circ}$ ({\it c.f.} planned values of $3.09$~m and $11.9^{\circ}$). 

Once this ``focus'' is established, the transmitter-to-receiver distance 
is fixed at 3.0~m and the oscilloscope trace of the received signal 
recorded for several fixed azimuthal angles spaced by $\approx 1^{\circ}$. 
The results of this procedure are shown in Fig.~\ref{timedepvsangle}. 
On comparing with Fig.~\ref{waveform}(a), it is clear that the broadcast 
signal (double peak, narrower then wider minimum) is only reproduced 
faithfully at an azimuthal angle of $11.6^{\circ}$ (orange, thicker curve). 
The time-dependent signals for angles $12.8^{\circ}$ and $10.5^{\circ}$ 
show distinct differences from the broadcast waveform; one only has to 
move a few more degrees away from $\Psi_0$ and any resemblance to the broadcast signal is lost.

This picture is confirmed by Fourier transforms of the oscilloscope data 
[Fig.~\ref{FTScramble}(a)]. At an angle of $11.6^{\circ}$ (red trace), 
the expected ``triangular’’ Fourier spectrum [{\it c.f.} Fig.~\ref{waveform}(b)] is produced. 
On moving $\approx \pm 2^{\circ}$ away, the relative amplitudes of the harmonics 
of 0.9~GHz change quite dramatically, showing that the frequency content present 
in the broadcast signal is being scrambled.

Measurements are then repeated at fixed transmitter-to-receiver distances 
either side of the target distance of $R_0 = 3.0$~m [Fig.~\ref{FTScramble}(b)-(d)]. 
Even azimuthal angles close to the target value (red traces) fail to yield the 
broadcast ``triangular’’ Fourier spectrum [compare with Figs.~\ref{FTScramble}(a), \ref{waveform}(b)], 
showing that the frequency content of the original broadcast signal is only 
reproduced when the distance {\it and} the azimuthal angle are close to the target values. 
Fourier transforms taken over wider angular ranges are given in the 
contour plots of Fig.~\ref{FTContour}, showing that the ``triangular’’ 
Fourier spectrum is not recovered as one moves farther from the target angle.
\begin{figure}[htbp]
	\centering
	\includegraphics [width=0.75\columnwidth]{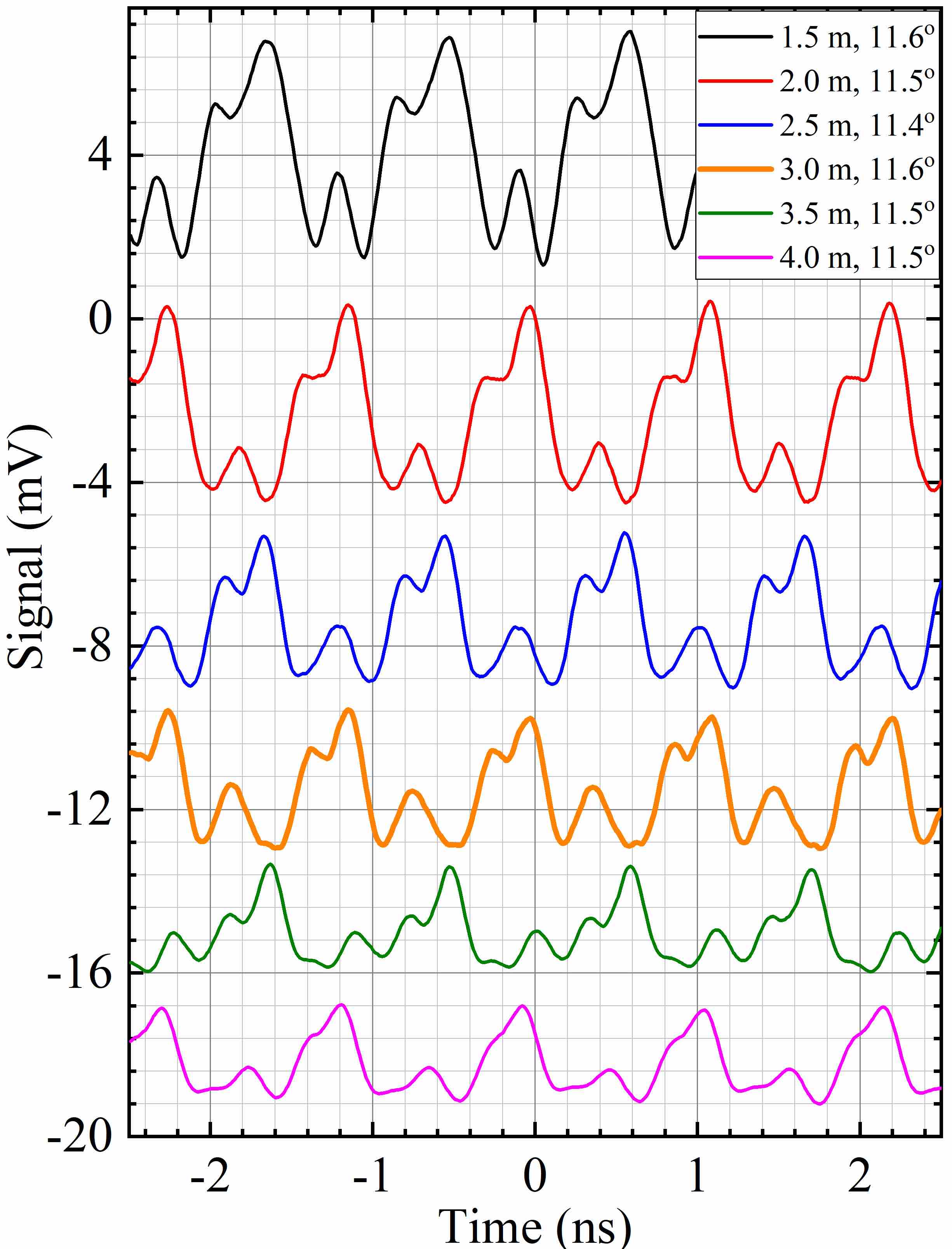}
	\caption{Time depndence of the signal received by the
		detector dipole for azimuthal angle close to $\Phi_0= 11.6^{\circ}$ and for 
		different antenna-to-detector distances shown in the key. The shape of the 
		transmitted waveform (Fig.~\ref{waveform}(b)) is only reproduced close to 
		the ``target'' $R_0=3.0$~m (orange trace).}
	\label{DistTimeDep}
\end{figure}
Fig.~\ref{DistTimeDep} shows the effect on the time dependence of the received signal 
caused by keeping the azimuthal angle close to $\Phi_0=11.6^{\circ}$ and varying the 
transmitter-to-receiver distance. Comparing Fig.~\ref{DistTimeDep} with Fig.~\ref{waveform}(a), 
it is clear that the broadcast signal’s time dependence (double peak, narrower and then 
wider minimum) is only reproduced faithfully at distances close to the target value of 
$3.0$~m (orange, thicker curve).

\section{Discussion}
\label{Rishi}
The data displayed in Figs.~\ref{waveform} to~\ref{DistTimeDep}
show that a continuous, linear, dielectric antenna in which a superluminal 
polarization-current distribution
accelerates can be used to transmit a broadband signal that is reproduced 
in a comprehensible form at a chosen target distance and angle;
as noted in the final paragraph of Section~\ref{accel},
effectively this signal is distributed onto a half circle~\cite{dissert} in
'the current implementation of the experiment~\cite{nichols}.
The requirement for this exact correspondence between broadcast and received signals
is that each point in the polarization-current distribution approaches the 
observer/detector at the speed of light at all times during its transit along the antenna.
This results in all of the radiation emitted from this point as it traverses the antenna
reaching the observer/detector at the same time [Fig.~\ref{ChirpFigure}(a)].
For other observer/detector positions, the time dependence of the signal is scrambled, due to the non-trivial
relationship between emission time and reception time [Figs,~\ref{ChirpFigure}(b), (c)]. 

\begin{figure}[htbp]
	\centering
	\includegraphics [width=0.99\columnwidth]{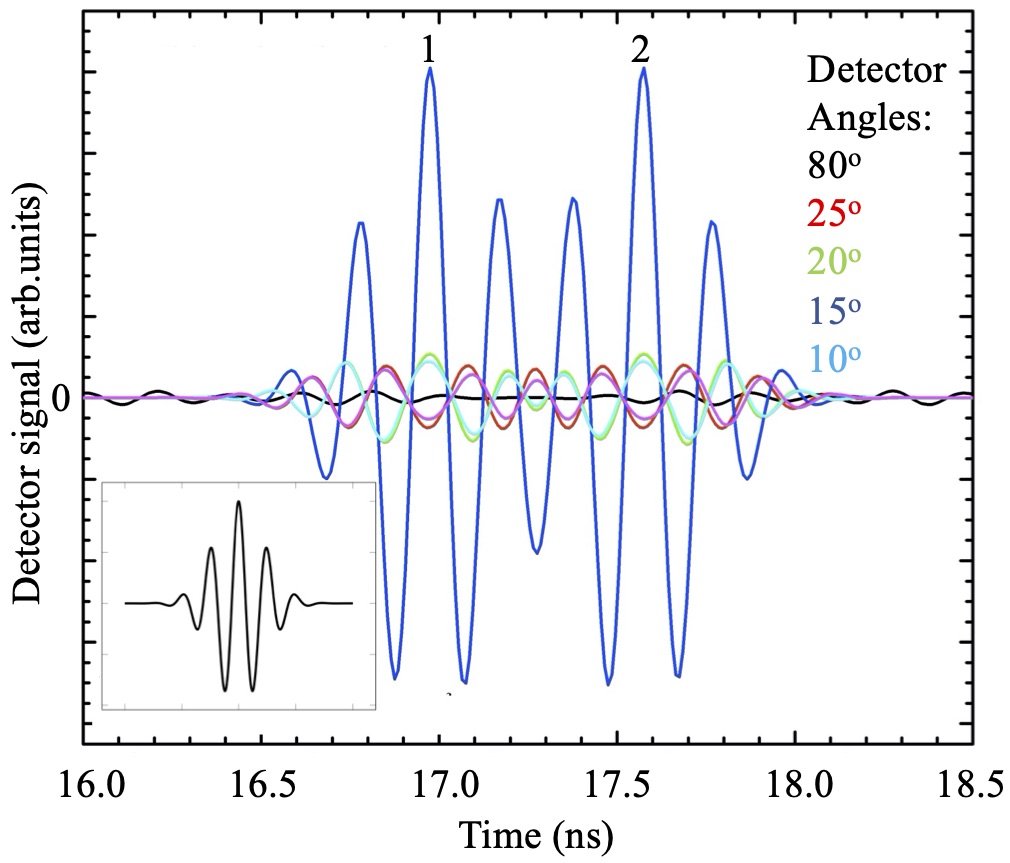}
	\caption{Simulation of a notional method for transmitting information
	only to a target point. The
	inset shows a ``bit'' (consisting of a Gaussian convoluted with a cosine)
	as it would appear in the time-dependence of the broadcast $E-$field
	[compare with Fig.~\ref{Concept}(c)].
	The main figure shows a calculation of the detected signal 
	at a range of 5~m. This results from 
	time-spacing two of these ``bits'' by three periods of the cosine function and 
	then subjecting them to the same acceleration scheme that is used to produce Fig.~\ref{ChirpFigure}.
	At the target angle of $15^{\circ}$ (dark blue), the ``bits'' (labelled 1 and 2) may be easily
	resolved. However, as soon as the detector is moved to other angles (labelled by the colors in the key),
	the received signal is much weaker and the ``bits'' become virtually impossible to distinguish.}
	\label{Extrablah}
\end{figure}

The primary r\^ole of the current paper is to introduce the above
effect and to demonstrate it experimentally. However, it is interesting
to suggest how a PCA might be employed to transmit signals that contain information.
Fig.~\ref{Extrablah} depicts a simulation of a simple version of such a concept. The inset
shows the time dependence of a wavepacket of launched $E-$field that could function as
a single ``bit''. Like the waveforms employed in
Figs.~\ref{Concept} and~\ref{ChirpFigure}, it consists of the convolution
of a Gaussian and a cosine. 
The main part of the Figure shows a calculation 
(using the techniques detailed in the SI~\cite{SI})
of the 
received signal due to the broadcast of two of these ``bits'',
spaced in time by three periods of the cosine function.
For ease of comparison, the antenna accelaration
scheme [{\it i.e.,} target angle $(15^{\circ})$ and distance (5.0~m)]
is the same as that employed in Fig.~\ref{ChirpFigure}.
At the target angle of $15^{\circ}$, the two ``bits'' can 
be distinguished clearly (labelled 1 and 2 in Fig.~\ref{Extrablah}); as one 
moves the receiver
away from the target angle
by as little as $5^{\circ}$, the received signal falls off in
amplitude and the individual ``bits'' become almost impossible to distinguish.
This example shows only two ``bits'';
however, a longer string of similar ``ones'' and ``zeros'' would also
suffer an analogous smearing as one moved away from the target position.

In this context, note that the {\it depth} of focus ({\it i.e.,} the range of 
distance and angle over which the signal is comprehensible)
depends strongly on the form and frequency content of the broadcast signal.
For example, the waveform used in the experiment, which encompasses 
frequencies from 0.9 to 4.5~GHz
[see Fig.~\ref{waveform}], results in a received signal that distorts 
relatively quickly as the detector moves out beyond the target 
distance of $R_0=3.0$~m at the target angle
[Fig.~\ref{DistTimeDep}]. 
By contrast, a relatively narrow-band broadcast signal [{\it e.g.,} Fig.~\ref{Concept}]
will be recognizable at the target angle over a wider range of detector distances~\cite{dissert}.
A full discussion of the criteria for the tightness  of ``information focusing'' 
demands detailed analysis of many different broadband signal types
and goes beyond the scope of the current work;
instead, it forms 
the basis of a subsequent paper~\cite{geology}.

This technique represents a contrast to conventional radio transmission methods.
In many instances of the latter, signals are broadcast with little or no directivity, selectivity 
of reception being achieved through the use of one or more narrow frequency bands~\cite{Balanis,Godara,adamy,ken}. 
In place of this, the current paper uses a spread of frequencies to transmit information 
to a particular location; the signal is weaker and has a scrambled time dependence elsewhere
[Fig.~\ref{ChirpFigure}].
A possible application may be in proposed 5G neighbourhood networks, where a single active 
antenna will sequentially spray bursts of information into a selection of target buildings 
around it~\cite{5G1,5G2}; ensuring that neighbours cannot easily understand what you are transmitting and 
receiving will be an important component. 

The work in this paper may also be relevant to pulsars, rotating neutron stars that possess 
very large, off-axis magnetic fields and plasma atmospheres~\cite{pulse1,pulse2}.
Pulsar periods of rotation $2\pi/\eta$ range from 1.5~ms to 8.5~s; a back-of-the-envelope calculation 
shows that at surprisingly small distances (85~km for the 1.5~ms pulsar; 40,000~km for the 8.5~s one)
from the rotation axis, the pulsar’s magnetic field will be travelling through 
its plasma atmosphere faster than the speed of light.
Hydrodynamical models of pulsars~\cite{hyd1,hyd2,hyd3} show the following:
(i)~electromagnetic disturbances (identifiable as polarization currents)
exist outside the light cylinder,
the orthogonal distance from the rotation axis $r_{\rm L}$ at which $\eta r_{\rm L} = c$;
(ii)~these disturbances rotate at the same angular velocity as the neutron star’s 
magnetic field (a requirement of Maxwell’s Equations), and so travel superluminally 
at radii outside the light cylinder; and 
(iii) the most intense disturbances are compact, in that they occupy a small fraction of 
the pulsar’s atmosphere.

For such a compact source, traveling on a circular path at 
faster-than-light speeds, a derivation given in the SI~\cite{SI} shows that 
a plot of observation/detection time $t_{\rm P}$ 
versus emission time $t$ exhibits ``plateaux'' [see Fig.~\ref{flatbits} of the SI]
at, and only at, a special polar angle determined by the source's tangential speed.
Apart from a single point at their center where ${\rm d}t_{\rm P}/{\rm d}t = 0$, 
these ``plateaux'' are not,  in fact,  flat~\cite{dissert}. 
However, there is a reasonable region of $t$ over which ${\rm d}t_{\rm P}/{\rm d}t \ll 1$,
so that a situation similar to that in Fig.~\ref{ChirpFigure}(a) may be possible.

Pulsars can potentially emit electromagnetic radiation via many 
mechanisms~\cite{pulse1,pulse2}, 
including thermal emission and other processes in their hot, plasma 
atmospheres, and dipole radiation from the rotating magnetic field of the 
neutron-star core; why then, might the pulsed radiation detected on Earth 
be dominated by the small volume of superluminal polarization current?
The similarity of the ``plateaux'' in Fig.~\ref{flatbits} of the SI to Fig.~\ref{ChirpFigure}(a)
provides an important clue. At the focus polar angle and over a short window
of $t_{\rm P}$, the frequency content of all of the 
emission processes occurring within the rotating polarization-current element will reproduce 
exactly, and result in a detected signal with greatly enhanced amplitude;
the result is similar to coherent emission~\cite{Brooker}, but via a completely
different mechanism.
At all other observation angles and observation times, 
radiation from the emission processes will 
superpose incoherently [{\it c.f.} Figs.~\ref{ChirpFigure}(b), (c)], 
leading to a greatly 
reduced amplitude, and scrambled frequency content.
The sharp focusing in the time domain at the focus polar angle
is likely to allow the radiation produced by the superluminal 
(outside the light cylinder) mechanisms to 
dominate the pulses.
Note that this explanation of the brightness of pulsar
pulses does not depend on the incorrect proposal~\cite{HADemolish}
of non-spherical decay advocated by Ardavan~\cite{HA1,HA2,HA3}.
\section{Summary}
The experiments in this paper show that a continuous, 
linear, dielectric antenna in which a superluminal 
polarization-current distribution
accelerates can be used to transmit a broadband signal that is reproduced 
in a comprehensible form at a chosen target distance and angle.
This is due to all of the radiation emitted from this point as it traverses the antenna
reaching the observer/detector at the same time.
For other observer/detector positions, the time dependence 
of the signal is scrambled, due to the non-trivial
relationship between emission (retarded) time and reception time .
The results may be relevant to 5G neighbourhood networks
and pulsar astronomy.
\vspace{-3mm}
\section*{Acknowledgments}
The experiments and calculations in this paper were supported by Los Alamos  
National Laboratory LDRD projects 20200285ER and 20180352ER.
We are grateful for additional support from
{\it LANL FY17 Pathfinder Fund Call for Technology Demonstration, PADGS:16-036}.
Much of this work was performed at the National High Magnetic Field 
Laboratory, USA, which is supported by NSF Cooperative Agreements 
DMR-1157490 and DMR-1644779, the State of Florida and U.S. DoE. 
J.S. acknowledges a Visiting Professorship from
the University of Oxford that enabled some of the 
calculations reported in this paper to be initiated. 
We thank Ward Patitz for his hospitality, assistance and very helpful suggestions
during antenna prototyping experiments carried out
at the FARM Range of Sandia National Laboratory.
We are grateful for a former collaboration with H and A Ardavan~\cite{jap,IEEE}
that gave the first hints of the potential of polarization-current antennas.


\vspace{3mm}
\noindent
$^\aleph$Contact emails: aschmidt@lanl.gov; jsingle@lanl.gov\\


\vspace{10mm}
\begin{Large}
\begin{center}
{\bf Appendix: Supplementary Information}
\end{center}
\end{Large}
\section{Demonstration of speed control of polarization currents}
\label{SIanimal}

\subsection{Superluminal speeds}

As an illustration, Sec.~2 of the main paper describes the simplest
method to produce a
polarization current moving at a constant speed~\cite{jap,IEEE,FeedMk1,FeedMk2,CryptoPatent}; 
the $j$th ($j = 1$, 2, 3....) element
of the antenna is supplied with time-dependent voltage differences \begin{equation}
V_j = [V_{\rm U}-V_{\rm L}]_j = V_0 \cos[\omega(t- j\Delta t)],
\label{volt1}
\end{equation}
where the symbols are defined in the main paper.
The voltages $V_j$ are usually $\leq 1$~V; under these, and much higher voltages,
alumina behaves as a linear dielectric,
so that the polarization {\bf P} in the $j$th element
will be proportional to the electric field~\cite{Jackson,Balanis,Bleaney}
generated by $V_j$.
The polarization current that
emits the radiation from the antenna is thus ``dragged along'' by the
time-dependent voltages applied to the elements
at a speed $v=a/\Delta t$,
 where $a$ is the separation of the centers of adjacent elements.

\begin{figure}[htbp]
\centering
\includegraphics [width=0.7\columnwidth]{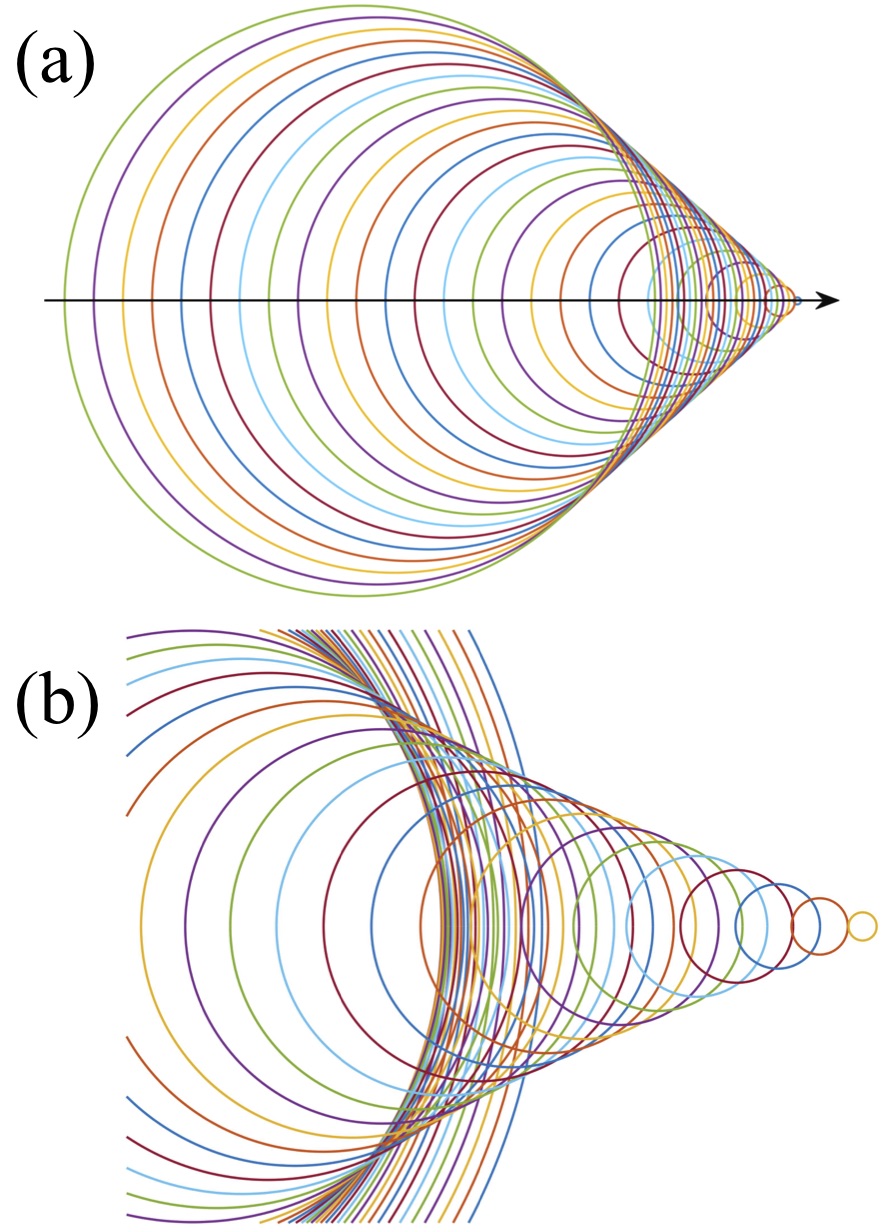}
\caption{(a)~Huygens wavelets emitted by a small source of electromagnetic radiation
traveling  along a rectilinear path with constant {\em superluminal} speed, 
$v/c = 1.5$. Since the source travels faster than the waves it emits,
it leads the advancing wave-front; the envelope of
the Huygens wavelets is a vacuum \v{C}erenkov cone
with half-angle $\sin^{-1}(c/v)$. 
(b)~Huygens wavelets emitted by a small source 
moving along a straight line with constant acceleration.
Having broken the ``light barrier'', the source
leaves a \v{C}erenkov envelope in its wake that has a slightly concave
lateral surface.
Note the clustering of many Huygens wavelets
to either side ({\it i.e.,} in a ring around the source's path);
this represents the arrival, in a relatively short period of observation time,
of radiation
emitted over an extended length of the source's 
path.
In the main paper, we optimize the
acceleration so that the Huygens wavelets emitted 
over the {\it entire} path of a point source
arrive simultaneously at a chosen ``target''.
[After Ref.~\onlinecite{dissert}, adapted from 
Schott~\cite{Schott}.]}
\label{SchottFig}
\end{figure}

\begin{figure}[htbp]
\centering
\includegraphics [width=0.85\columnwidth]{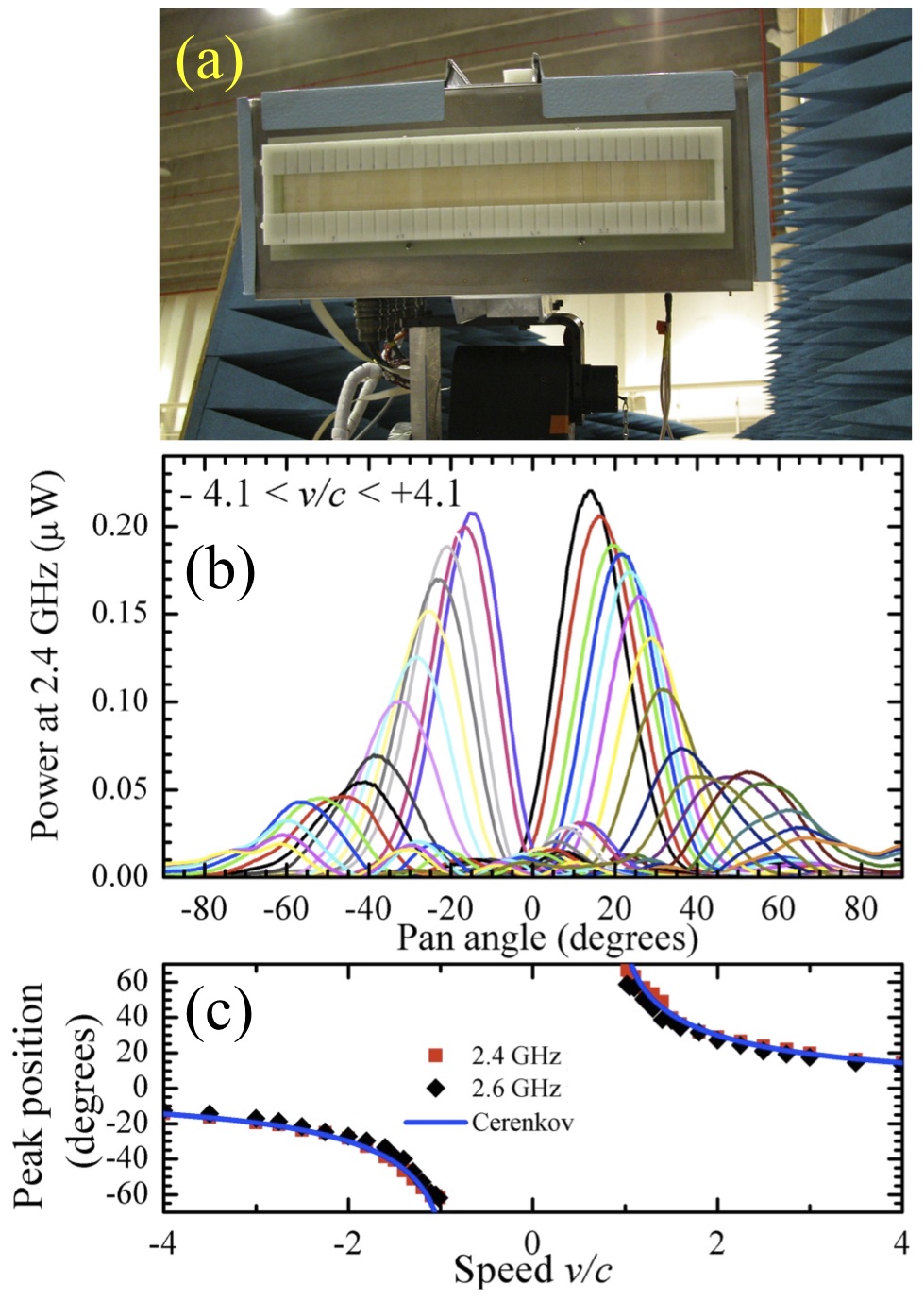}
\caption{(a)~The antenna used in the vacuum \v{C}erenkov
demonstration placed on a stepper-motor-driven
pan/tilt mount.
(b)~Power received from the antenna
run at several different constant speeds $v$, ranging from
$-4.1c$ to $-1.1c$ and $+1.1c$ to $+4.1c$, 
versus azimuthal (pan) angle. The emitted frequency 
is $\frac{\omega}{2 \pi} = 2.4$~GHz and the detector is 
5.36~m from the antenna in an anechoic chamber.
(c)~Peak power
angle versus polarization-current 
speed $v$ with $\frac{\omega}{2\pi} =2.4$~GHz
and $\frac{\omega}{2\pi} = 2.6$~GHz (points); the curve is the
prediction for the vacuum \v{C}erenkov effect
[Eq.~\ref{coughtheory}].}
\label{speedvary}
\end{figure}

In the early years of the 20$^{\rm th}$ Century, both
Sommerfeldt~\cite{Sommerfeldt} and Schott~\cite{Schott} showed that 
emission of electromagnetic radiation
from such a moving source can only occur when $v>c$, the speed of light {\it in vacuo}.
Schott demonstrated~\cite{Schott} that the Huygens wavelets from each point
in the moving polarization current form a conical envelope
with aperture $\sin^{-1}(c/v)$ [Fig.~\ref{SchottFig}(a)].
Translating this to an extended, moving polarization current that fills the entire antenna, this results
in emitted power that should peak at an azimuthal angle
\begin{equation}
\phi = \sin^{-1}\left(c/v \right).
\label{coughtheory}
\end{equation}
To demonstrate this effect,
we use data from the antenna~\cite{FeedMk1} shown in Fig.~\ref{speedvary}(a).
Like the antenna in the main paper, it comprises 32 elements
and employs alumina $(\epsilon_{\rm r} \approx 10)$ as a dielectric, 
but it has a smaller element spacing of
$a=10.87$~mm.
The detected power is monitored using a Schwarzbeck Mess 
calibrated dipole
whilst the antenna is rotated on a turntable to vary the 
azimuthal angle $\phi$.

Fig.~\ref{speedvary}(b) shows
detected power (in $\mu$W) versus $\phi$
for the antenna running 
at a series of constant speeds $v$, set by varying $\Delta t$ in Eq.~\ref{volt1}.
When plotted in these linear power units,
the azimuthal dependence is clearly dominated by a single, large peak,
the angle of which depends on $v$.
Fig.~\ref{speedvary}(c)
shows that the azimuthal angle $\phi$ at which peak power occurs
varies with $v$
as expected for the vacuum \v{C}erenkov 
effect~\cite{Schott} [Eq.~\ref{coughtheory}].

\subsection{Subluminal speeds}
\begin{figure}[tbp]
	\centering
	\includegraphics [height=12.5cm]{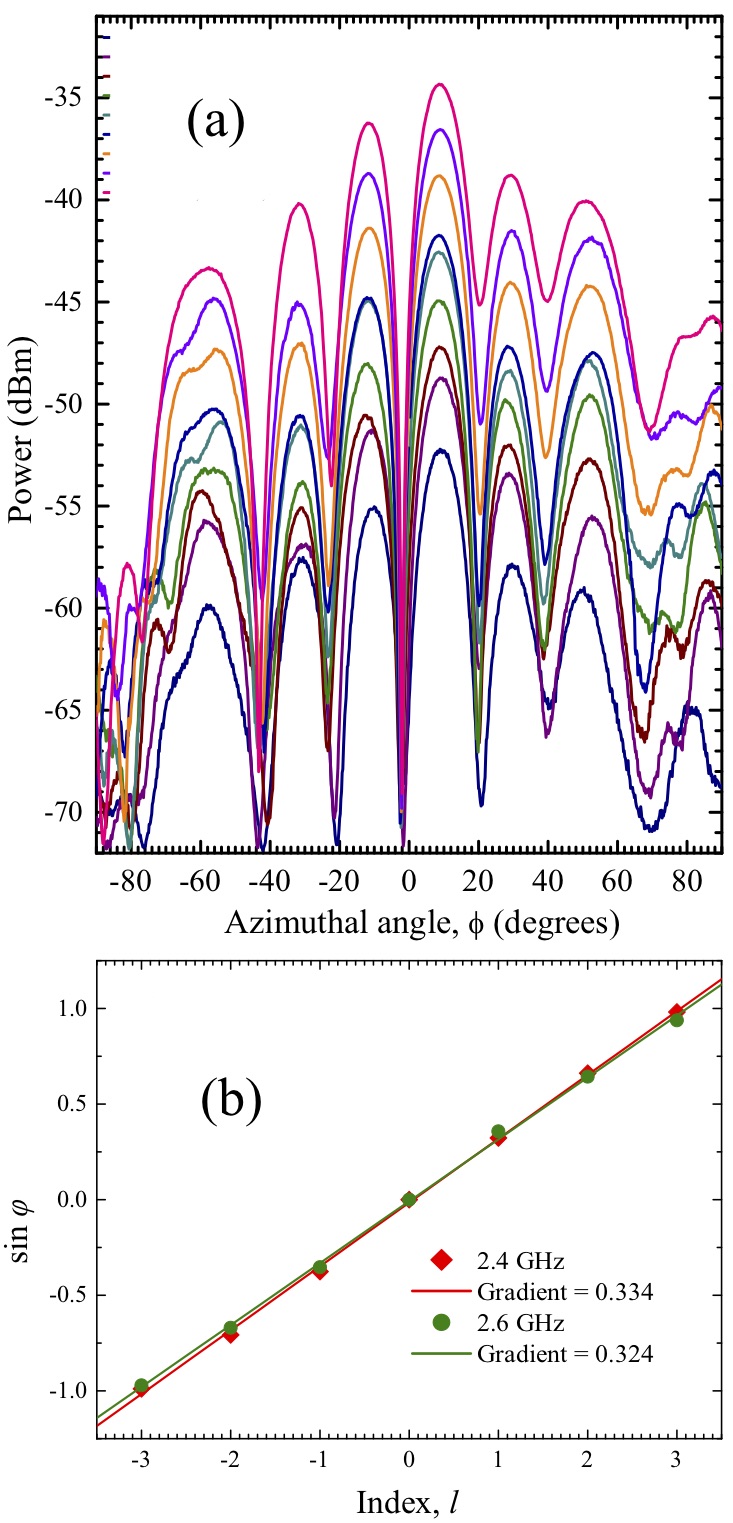}
	\caption{(a)~Received power versus azimuthal
		angle $\phi$ for antenna-to-detector distances 0.95, 1.4, 1.8, 2.3, 3.2, 4.1, 5.2, 5.4
		and $8.6$~m, and subluminal source speed $v/c = 0.5$; 
		the antenna used is that in Fig.~\ref{speedvary}(a)
		and data are shown for $\omega/2\pi = 2.6$~GHz.
		(b)~A plot of $\sin \phi$ versus index $l$, where $\phi$ denotes
		the angle at which a minimum in power occurs,
		for a distance of 8.6~m [see (a)].
		Data are points, and the lines are straight-line fits,
		giving gradients of 0.33 and 0.32 for
		$\omega/2\pi = 2.4~$GHz and 2.6~GHz points respectively.}  
	\label{lin0p5}
\end{figure}

When polarization-current antennas are run at constant speeds
$v/c < 1$, the power detected oscillates
as a function of azimuthal angle. 
Fig.~\ref{lin0p5}(a) gives an example of this behavior;
to obtain these data, the antenna shown in Fig.~\ref{speedvary}(a)
was run at $v/c = 0.5$.
In addition to the oscillations, the peak power
is some $15-20$~dB lower than for speeds
$v/c\approx 2-4$.

The power oscillations in Fig.~\ref{lin0p5}(a)
are similar to those
from a two-slit diffraction experiment in which
the light from one slit is $(2\aleph +1)\pi$ out of phase with 
that from the other, where $\aleph$ is an integer.
In the far field, such a two-slit experiment would give
minima that occur when~\cite{Brooker} 
\begin{equation}
b\sin \phi = l \lambda,
\label{young}
\end{equation}
where $b$ is the spacing of the slits, $l$ is an integer
and $\lambda$ is the wavelength.
Hence, a plot of $\sin \phi$ versus $l$ should be a straight
line, with gradient $\lambda/b$.
Fig.~\ref{lin0p5} shows that the minima
indeed obey this relationship. The reasons for this
will be discussed in more detail below.

\section{Feeds to antenna elements}
\begin{figure}[tbp]
	\centering
	\includegraphics [height=11cm]{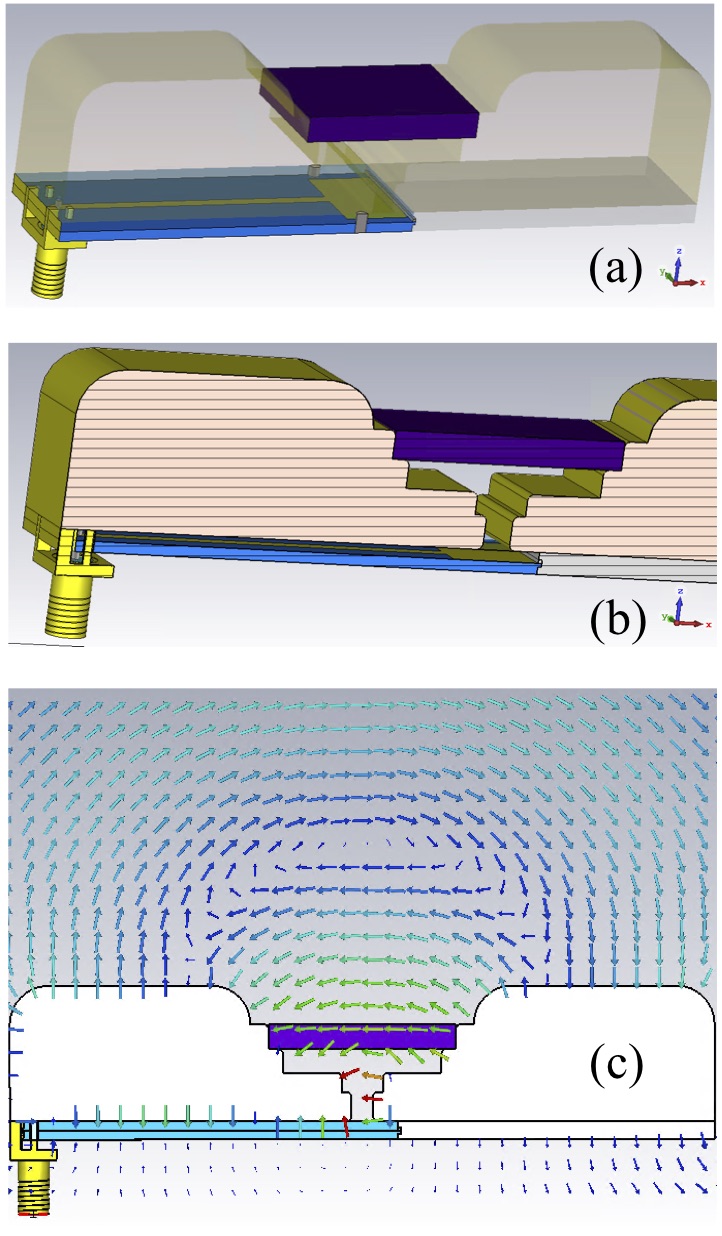}
	\caption{(a, b)~Two views of an individual element of the antenna shown in Fig.~2 of the main paper.
	The RF signal is fed from an SMA connector and coupled to a micro-stripline (gold on blue). 
	The final rectangular patch underneath the cut-out opening and the shape of the stepped opening above transform 
	the field pattern into a linearly polarized electric field across the alumina dielectric (purple) between the electrodes. 
	The whole upper part of the element is made up from two pieces of copper-plated G10. 
	(c)~The fields (arrows) from the polarization current in the element radiate out of the top surface only. 
	Transverse components of the fields are suppressed by the proximity of the neighboring antenna elements. 
	}  
	\label{ElFig}
\end{figure}
An important ingredient of the design of polarization-current antennas is the 
way in which radiofrequency (RF) signals are transformed from cylindrical
propagating modes in $50~\Omega$ coaxial cable to a linear electric field
applied by the electrodes to the dielectric~\cite{FeedMk1,FeedMk2, dissert}.
Figs.~\ref{ElFig}(a,b) show how this is achieved in the antenna depicted in Fig.~2 of the main paper.
Most of the antenna element is made from G10 composite, with the upper surface plated with copper.
The radiofrequency signal is fed from an SMA connector ($50~\Omega$) and coupled to an
impedance-matched micro-stripline built on
standard circuit board. The final transformation into the desired, linear field pattern 
to be applied to the dielectric
is achieved by a rectangular patch underneath the 
cut-out opening and the shape of the stepped opening above.

Note that the individual elements cannot be designed, modelled or tested on their own;
isolated, they do not have the desired characteristics for the function of the antenna.
Instead, their performance depends on the presence of neighboring elements.
Fig.~\ref{ElFig}(c) shows how the fields from the polarization current in the element 
radiate out of the top surface only; the transverse components of the fields 
are suppressed by the proximity of the adjacent antenna elements.
The design of the elements also
suppresses radiation out of the back of the antenna,
increasing the directivity.

In practice, the antenna elements can be built individually and then
combined into different configurations optimized for particular applications~\cite{dissert}.
Alternatively, all of the elements for an antenna can be fashioned on a single monolithic substrate
(machined from a G10 block by a CNC mill)
for strength and rigidity.
\section{Calculation of emitted radiation}
\label{SIaccel}
\subsection{Basic principles}
Simulated antenna emissions are used to illustrate the principles of the experiment 
in the main body of the paper.
Therefore, a brief explanation of the simulations
and a validation via comparison with experimental data
taken using the antenna shown in Fig.~\ref{speedvary}(a)
are now given.

Despite the discrete nature of the electrodes, simulations of 
our antennas performed with off-the-shelf electromagnetic 
software packages show that fringing fields of adjacent 
electrode pairs lead to a voltage phase that 
varies under the electrode~\cite{frank}; 
{\it i.e.,} the phase varies much more smoothly than the 
discrete electrodes suggest.
Therefore we represent the position dependence of
the voltage applied across the dielectric
as a continuous function, giving two examples below.
\\

\noindent
{\bf (i)~For a constant speed} ({\it c.f. }Eq.~\ref{volt1}) we 
consider a traveling, oscillatory voltage applied symmetrically
across the dielectric in the vertical $z$ direction,
$V = V_0{\rm e}^{{\rm i}(\omega t - ky)}$.
Here $y$ is the distance along
the antenna's long axis and $V_0$ and $k$ are constants.
Let the dielectric extend from $z=-z_0$ to $z=+z_0$ in the vertical direction; 
assuming that it is uniform, 
the potential at a general position $(y,z)$ in the dielectric is
\begin{equation}
V(y,z,t)=V_0\frac{z}{2z_0}{\rm e}^{{\rm i}(\omega t - ky)}.
\end{equation}

\noindent
{\bf (ii)~For the wavepacket} shown in Fig.~2, main 
paper, the voltage is given by
$V_0 {\rm e}^{{\rm i}\omega [t-p(y)]}{\rm e}^{-\alpha^2[t-p(y)]^2},$ 
where $\alpha$ is a constant.
This consists of a Gaussian convoluted with a travelling wave;
both have the $(y,t)$ dependence [given by $p(y)$] required for the motion described 
in the main paper.
Under the same assumptions as (i), the potential at $(y,z)$ in the dielectric is
\begin{equation}
V(y,z,t) = V_0\frac{z}{2z_0} {\rm e}^{{\rm i}\omega [t-p(y)]}{\rm e}^{-\alpha^2[t-p(y)]^2}.
\label{blobblob}
\end{equation}

For either equation, the polarization {\bf P} is obtained~\cite{Bleaney}
by
\begin{equation}
{\bf P} = \epsilon_0(\epsilon_r-1){\bf E} = \epsilon_0(\epsilon_r-1)[-\nabla V(y,z,t)].
\end{equation}
Differentiating with respect to time,
we obtain a polarization-current density~\cite{Bleaney}
\begin{equation}
{\bf J}(y,z,t)=\frac{\partial {\bf P}}{\partial t} 
\label{boristime}
\end{equation}
	
In evaluating the emitted radiation, we consider only the
contribution of ${\bf J}$ in the dielectric;
there is a negligible conduction current,
and we neglect the free charges that exist
only at the interface between the dielectric and the electrodes.
In this situation, the following equation~\cite{Balanis,Bleaney,Jefimenko} 
can be used to evaluate the
magnetic vector potential ${\bf A}$ at the 
observer/detector's remote location
${\bf r}_{\rm P} = (X,Y,Z)$ and at
the observation time $t_{\rm P}$:
\begin{equation}
{\bf A}({\bf r}_{\rm P},t_{\rm P}) = \frac{\mu_0}{4\pi}\int^{z_0}_{-z_0}\int^{y_0}_{-y_0}\int^{x_0}_{-x_0}\frac{{\bf J}({\bf r},t)}{|{\bf r}_{\rm P} - {\bf r}|}{\rm d}x {\rm d}y {\rm d}z
\label{eq:vectorPotential}
\end{equation}
Here, 
${\bf r}=(x,y,z)$ is a coordinate
within the dielectric.
The integration is carried out over the volume of the 
dielectric; 
as in the  main paper, its length is $2y_0$
and its thickness in the $x$ direction is $2x_0$.
The retarded time $t$ varies for different locations {\bf r}
within the dielectric:
\begin{equation}\label{eq:time}
t = t_{\rm P} - \frac{|{\bf r}_{\rm P}-{\bf r}|}{c}.
\end{equation}
Here $c$ again denotes the speed of electromagnetic waves in the 
medium (assumed to be uniform) 
between the source and the observer.

\begin{figure}[tbp]
	\centering
	\includegraphics[height=6.6cm]{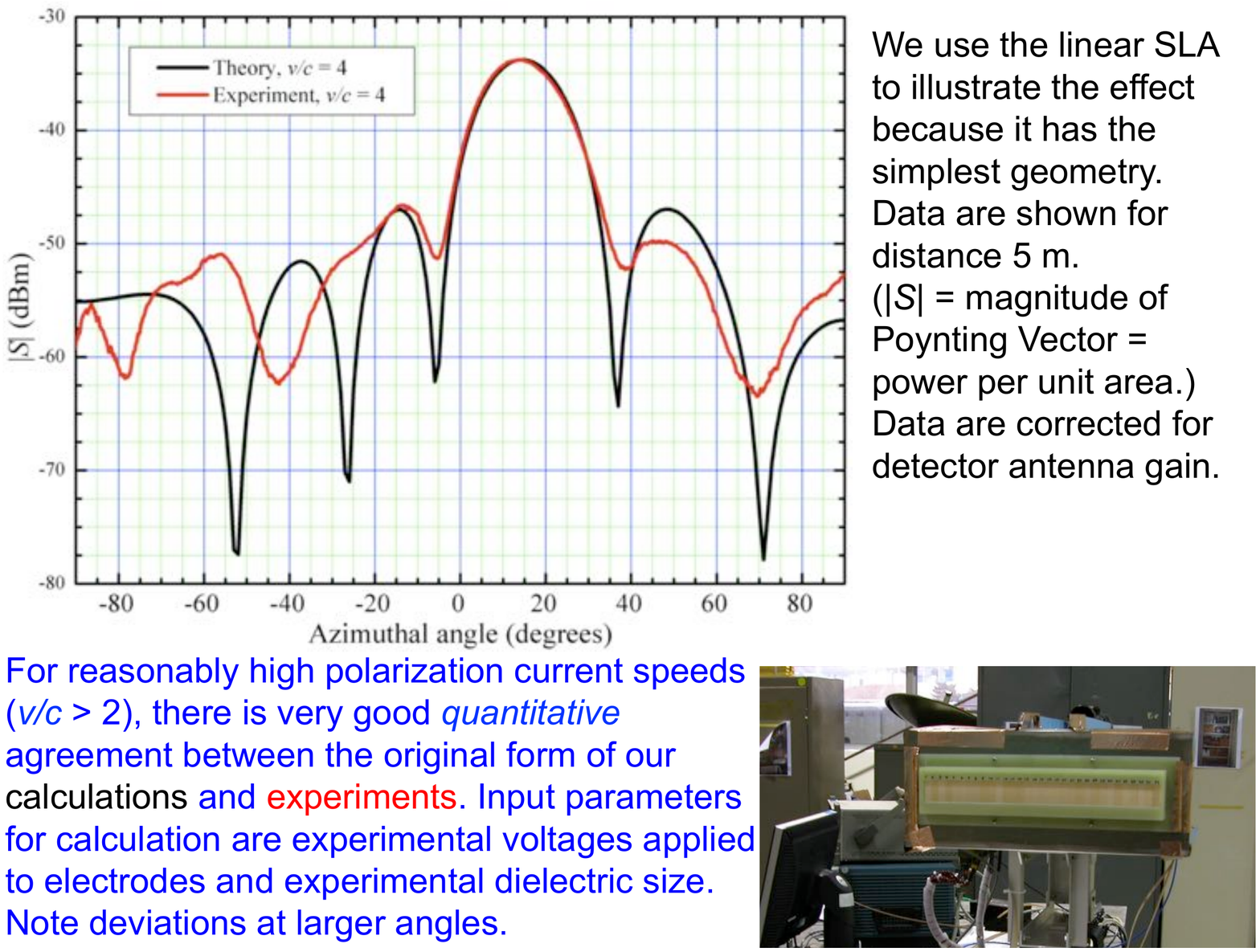}
	\caption{Modulus of Poynting vector $|S|$ [Eq.~\ref{poynty}] versus azimuthal
		angle $\phi$ (black) predicted by the calculations.
		The antenna dimensions are based on
		the example shown in Fig.~\ref{speedvary}(a).
		The frequency is $\omega/2\pi = 2.4$~GHz, $v/c = 4$ and the 
		antenna to detector distance is 5~m. 
		Experimental data for the same antenna,
		measured at the same distance (5~m) in
		an anechoic chamber 
		are shown in red. After
		correction for the gain of the
		detector dipole and losses in its cables, there
		is a reasonable quantitative
		match between experiment and calculation.}  
	\label{JeremyCorbyn}
\end{figure}

The corresponding radiation fields are derived from
differentiating ${\bf A}$ with respect to the
{\it observer's} coordinates $(X,Y,Z,t_{\rm P})$.
The electric field is given by~\cite{Jackson,Bleaney}
\begin{equation}
{\bf E}(X,Y,Z,t_{\rm P}) = \frac{\partial {\bf A}(X,Y,Z,t_{\rm P})}{\partial t_{\rm P}},
\label{eq:E-field}
\end{equation}
and since the magnetic flux is always solenoidal
({\it i.e.,} $\nabla\cdot{\bf B} = 0$), 
the magnetic flux density ${\bf B}$ is given by~\cite{Jackson,Bleaney}
\begin{equation}
{\bf B}(X,Y,Z,t_{\rm P}) \equiv \nabla\times{\bf A}(X,Y,Z,t_{\rm P}).
\label{eq:B-field}
\end{equation}
We again emphasize that the 
curl operator $(\nabla\times)$ employs 
{\it observer} coordinates $(X,Y,Z)$.
In free space, the magnetic field is simply ${\bf H} =\frac{1}{\mu_0}{\bf B}$.
The received power is computed by evaluating the Poynting vector~\cite{Balanis,Bleaney}
\begin{equation}
{\bf S}(X,Y,Z,t_{\rm P})={\bf E}(X,Y,Z,t_{\rm P})\times {\bf H}(X,Y,Z,t_{\rm P}).
\label{poynty}
\end{equation} 
The steps up to and including Eq.~\ref{boristime} are  carried out analytically; the integral [Eq.~\ref{eq:vectorPotential}], the two derivatives [Eqs.~\ref{eq:E-field} and \ref{eq:B-field}]
and their cross-product [Eq.~\ref{poynty}]
are evaluated numerically~\cite{dissert}.

\begin{figure}[htbp]
	\centering
	\includegraphics[height=11.0cm]{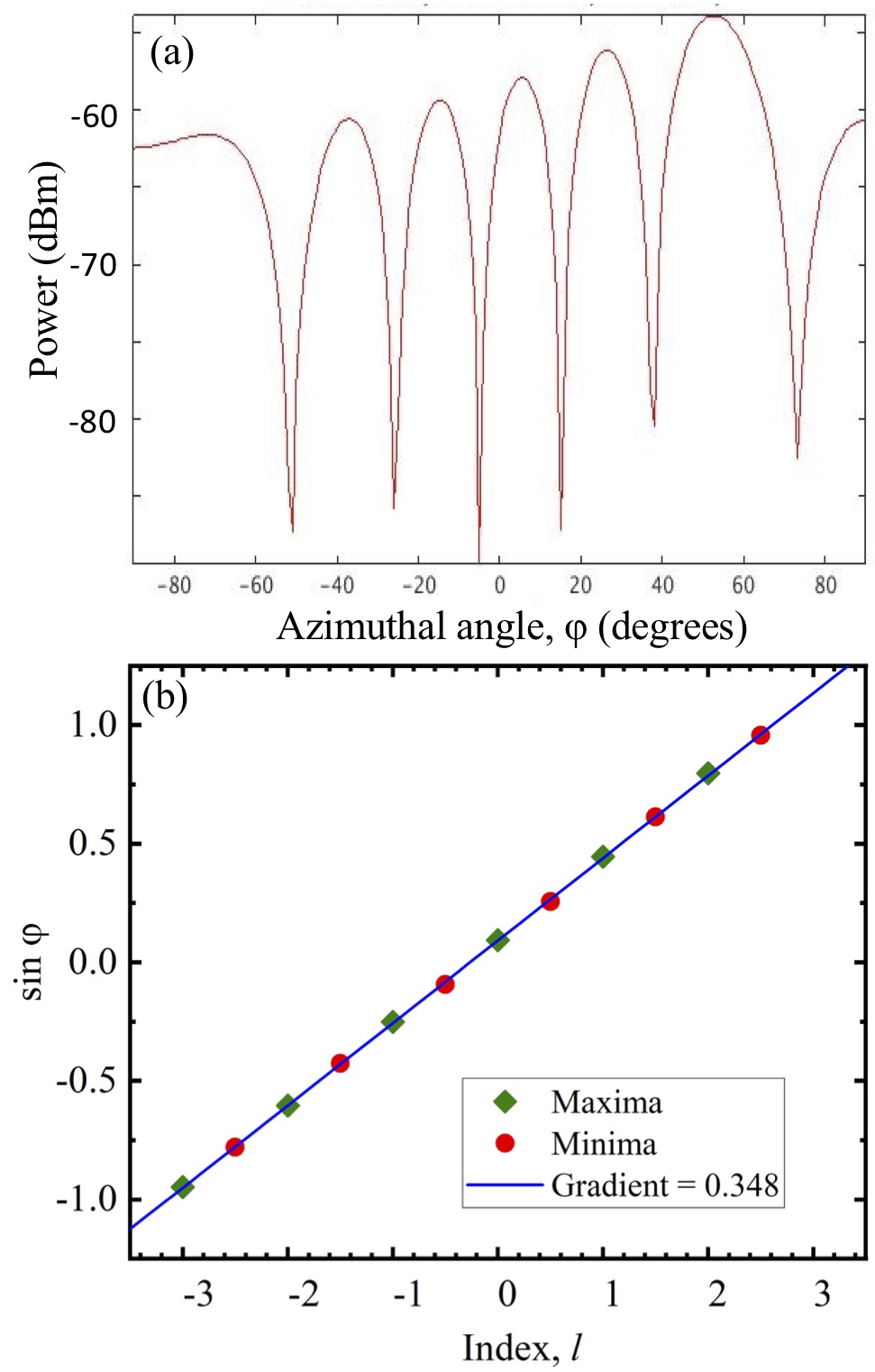}
	\caption{a)~Simulated received power versus azimuthal
		angle $\phi$ for an antenna-to-detector distance of 10.0~m, 
		and subluminal source speed $v/c = 0.5$; 
		the antenna simulated is that in Fig.~\ref{speedvary}(a)
		and data are shown for $\omega/2\pi = 2.5$~GHz.
		As is the case for the experimental data shown in
		Fig.~\ref{lin0p5}(a), the simulated power oscillates
		with azimuthal angle.
		(b)~A plot of $\sin \phi$ versus index $l$, where $\phi$ denotes
		the angle at which a maximum (blue) or minimum (red) in power occurs,
		for the numerical data in (a). The gradient of the fitted
		line is 0.35, very close to the experimental values
		shown in Fig.~\ref{lin0p5}(b).
	}
	\label{minima}
\end{figure}

\subsection{Numerical results}
\label{HenPower}
An example of the numerical calculations for constant $v/c$
is compared with experimental data in Fig.~\ref{JeremyCorbyn}.
The experimental conditions [voltage applied
to the electrodes, polarization-current speed $(v/c =4)$,
dielectric dimensions- the antenna is that shown in Fig.~\ref{speedvary}(a)]
are used as model input parameters.
After correction for the gain of the receiver dipole and cabling used
in the experiment, there is a reasonable quantitative
match between
data and theory, especially close to the main lobe.
At larger angles, the match is less good.
This is understandable, because the
subsidiary minima are very dependent on the
precise phases of the signals applied to each antenna element,
which are subject to errors of a few degrees in the experiment~\cite{dissert}. 
Similar quantitative agreement between model and anechoic chamber data
was obtained for all speeds above $v/c = 2$. 

Turning to subluminal speeds, 
Fig.~\ref{minima}(a) shows a simulation
of the antenna shown in Fig~\ref{speedvary}(a)
run at $v/c=0.5$, frequency $\omega/2\pi=2.5$~GHz 
and for an antenna-to-detector distance of 10~m.
The calculated power oscillates with azimuthal angle
in a similar manner to the experimental data 
[see Fig.~\ref{lin0p5}(a)].
Fig.~\ref{minima}(b) plots the
$\sin \phi$ versus index $l$ for the model calculation
shown in Fig.~\ref{minima}(a); positions of both minima
and maxima are shown. 
The gradient of the fitted
line is 0.35, close to the experimental 
gradients shown in Fig.~\ref{lin0p5}(b).

The results in Figs.~\ref{lin0p5} [experimental]
and \ref{minima} [model] may be understood as follows~\cite{dissert}.
Ideally, {\it no} vacuum \v{C}erenkov radiation should be emitted
for $|v/c| <1$, as the emission angle
$\sin^{-1}c/v$ becomes imaginary~\cite{Schott,Ginzburg}.
In this context, ``ideal'' implies an infinitely-long
source of identical elements, 
in which the radiation from all elements superposes
to produce no net emission.
However, the 32-element antenna shown in Fig.~\ref{speedvary}(a) is of finite
length, so that the radiation measured in the $v/c = 0.5$
experiments is likely to come mostly from the {\it ends} of the array;
the elements at the ends have adjacent elements on only one side.
Hence, one might expect that
about half of their emitted power would be cancelled out,
so that the two end elements behave like
a double-slit experiment emitting a total
power $\sim (1/2)\times 2\times(1/32) =1/32$ of the total power of the array.
Converting into dB, $10\log_{10}(1/32)=-15$~dB,
explaining why the peak subluminal emission
in both experiment and simulations
is $15-20$~dB lower than the peak vacuum \v{C}erenkov
power produced at speeds $v/c>2$,
where all 32 elements contribute.

Using Eq.~\ref{young} with the relevant wavelengths $\lambda = 2\pi c/\omega$,
the gradients of the experimental data for $\omega/2\pi = 2.4$~GHz and 2.6~GHz
[Fig.~\ref{lin0p5}(b)]
yield $b=370$~mm and $b=360$~mm, and the simulation [$\omega/2\pi=2.5$~GHz; Fig.~\ref{minima}(b)]
$b=347$~mm, values that are very close
to the 348~mm overall length of the array of 32 elements [Fig.~\ref{speedvary}(a)].
Moreover, for $v/c=0.5$, the phase difference
of elements $j=1$ and $j=32$ is very close
to $11\pi$, explaining why the experimental ``interference pattern''
has {\it minima} quite close to the values of $\phi$ given by Eq.~\ref{young}~\cite{dissert}.

Examples of simulated emission from propagating pulses
produced by driving voltages
analogous to that given in
Eq~\ref{blobblob} are displayed in the main body of the paper.

\section{Small region of polarization current in Superluminal Rotation}\label{S:SuperluminalRotation}
\begin{figure}[htbp]
	\centering
	\includegraphics [width=0.6\columnwidth]{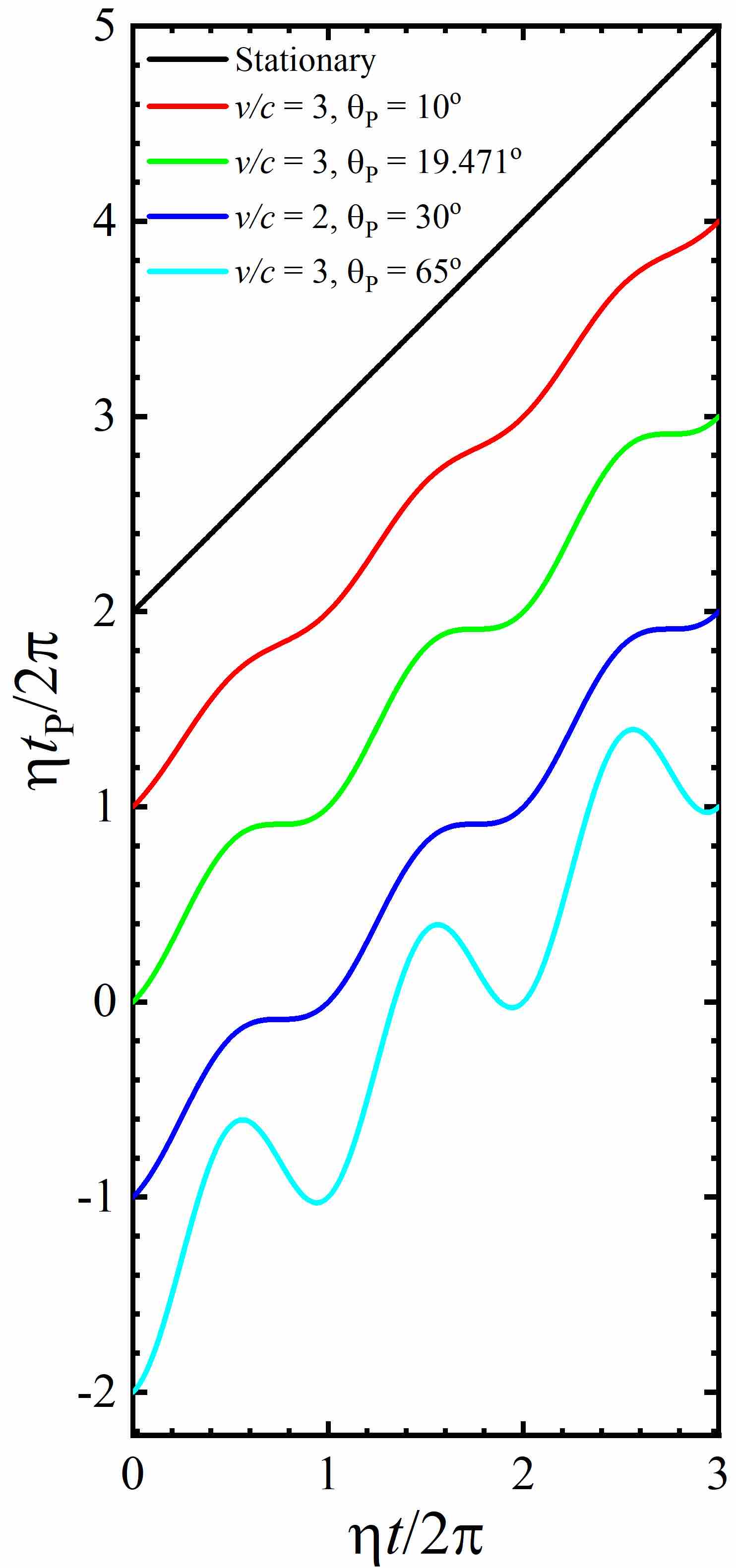}
\caption{Various forms that Eq.~\ref{C:3} can take
depending on the position of the observer (refer to Fig.~\ref{critical});
the times on both axes are scaled by the source angular
velocity, $\eta$.
The curves are shown for $r_{\rm P}= 100r_{\rm L}$,
where $r_{\rm L} = c/\eta$ is the light-cylinder radius,
and $\varphi_{\rm P}=0$; finite $\varphi_{\rm P}$ merely introduces a phase shift.
The observation polar angles $\theta_{\rm P}$ and the source
tangential speed $v$ are shown in the inset key.
Note that $\theta_{\rm P} = \sin^{-1} (c/v)$
for the green and dark blue curves.
Curves are offset by vertical increments of
$\eta t_{\rm P}/2\pi =1$ for clarity,
and, for reference, the black line shows the $1:1$
correspondence between emission
and observation time that would occur
for a stationary source.
(Curves similar to those shown in light blue and red
were first derived in Ref.~\onlinecite{Schott};
curves analogous to those shown in green and
dark blue are presented in Ref.~\onlinecite{HA1}.)
}
	\label{flatbits}
\end{figure}
\begin{figure}[h!]
	\centering
	\includegraphics[height=2.5in]{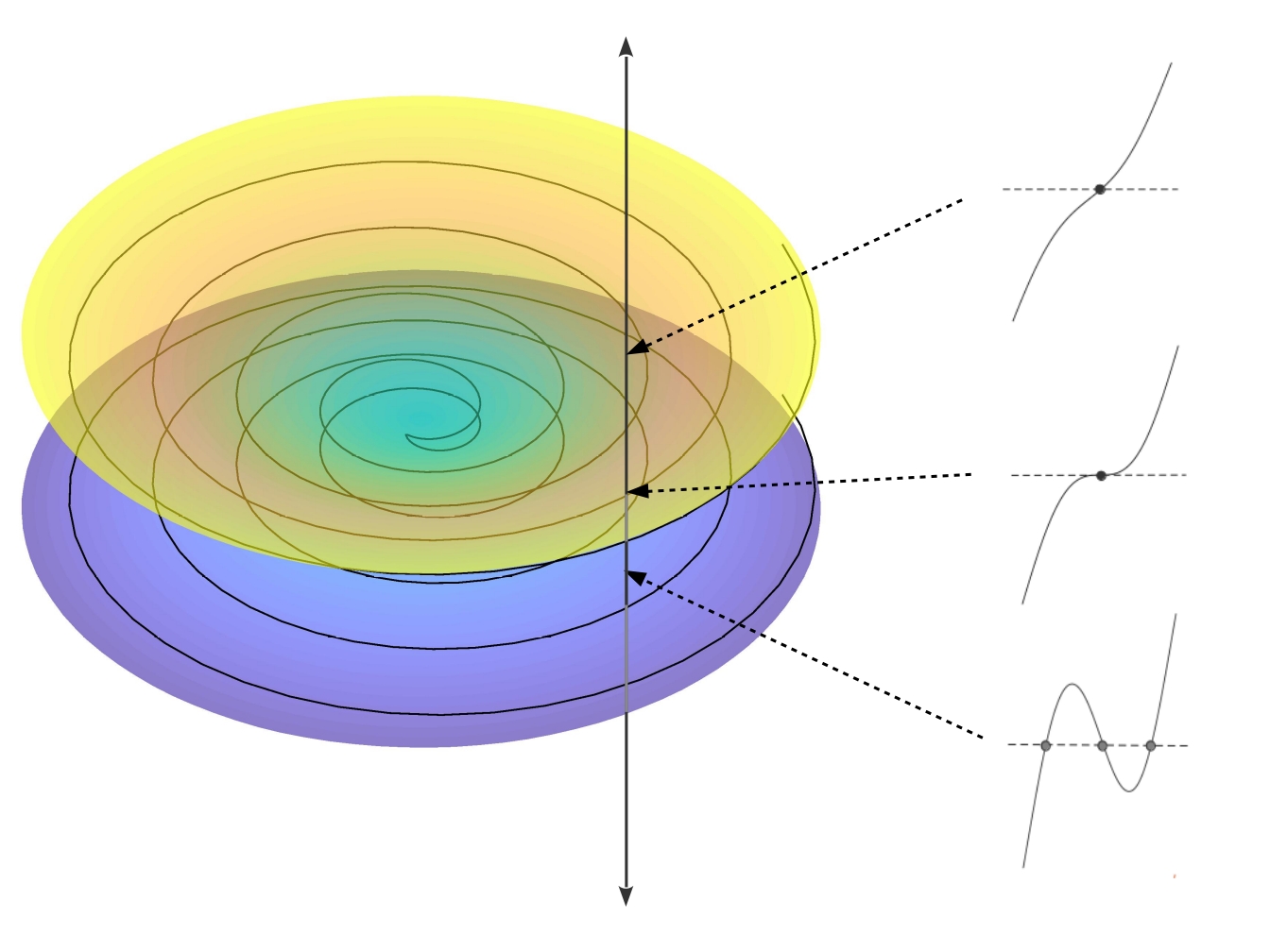}
	\caption{
Illustrations of the regions in which Eq.~\ref{C:3} takes the various forms 
shown in Fig~\ref{flatbits}. The $z$ axis is vertical,
and the circular source orbit lies in the green $(z=0$) plane.
The ``double-funnel’’ structure (yellow and purple; symmetrical
about the rotation plane $z=0$) represents observer
positions at which the source momentarily approaches
with an instantaneous speed of $c$ and with zero acceleration.
Instantaneous positions of points of this nature are shown as the ``spiralling''
fine line superimposed on the surface.
The shape of Eq.~\ref{C:3} for each region is shown by the curves plotted around the figure, 
with arrows linking each curve to the place in which it would be observed. 
The central arrow point lies exactly on the double-funnel structure;
the black vertical line traverses the surfaces and is
included to clarify the positions of the arrow points. [After Ref.~\onlinecite{dissert}.]}
\label{critical}
\end{figure}

Consider a polarization-current element of small volume that 
rotates in the $xy$-plane at radius $r$ with angular velocity $\eta$
and emits radiation (hereafter referred to as {\it the source}).
In terms of the cylindrical coordinates $r$, $\varphi$ and $z$, the path ${\bf r}(t)=(r,\varphi,z)$ of the source 
is given by
\begin{equation} \label{C:1}
r={\rm const},~~~~\varphi=\varphi_Z+\eta t, ~~~~
z=0,
\end{equation} 
where the coordinate $\varphi_Z$ denotes the initial azimuthal position of $\varphi$ 
and is, without loss of generality, assumed to be zero from now on. 
The wave fronts that are emitted by this point source in an empty and 
unbounded space can then be described by
\begin{equation}\label{C:1a}
|{\bf r}_{\rm P}-{\bf r}|=c(t_{\rm P}-t);
\end{equation}
as before, the constant $c$ denotes the wave speed and the observation/detection 
point is defined as $({\bf r}_{\rm P}, t_{\rm P})=(r_{\rm P},\varphi_{\rm P},z_{\rm P},t_{\rm P})$. 
Inserting \eqref{C:1} into \eqref{C:1a} and utilizing the theorem of Pythagoras we find 
that the distance $R$ which separates the source from the observer/detector is given by 
\begin{equation}
R(t) = \left[
z_{\rm P}^2+r_{\rm P}^2+r^2-2rr_{\rm P}\cos(\varphi_{\rm P}-\eta t)
\right] ^\frac{1}{2}\,.
\label{RtyFarty}
\end{equation}
In consequence, the relationship between the emission time $t$ and the reception (detection) time
$t_{\rm P}$ must satisfy
\begin{align}\label{C:3}
t_{\rm P}  &= t+\frac{R(t)}{c}\\
\notag
&= t+\frac{1}{c}\left[
z_{\rm P}^2+r_{\rm P}^2+r^2-2rr_{\rm P} \cos(\varphi_{\rm P}-\eta t)
\right] ^\frac{1}{2}.
\end{align}
Fig.~\ref{flatbits} shows the three generic forms that Eq.~\ref{C:3} can take, whilst
Fig.~\ref{critical} identifies the observer positions corresponding to these forms.
In the following discussion, it is helpful to define the observer's polar angle as
\begin{equation}
\theta_{\rm P} =\tan^{-1}\left(\frac{r_{\rm P}}{z_{\rm P}}\right).
\end{equation}
First, the green and dark-blue curves in Fig.~\ref{flatbits}
correspond to an observer positioned on (or very close indeed to) the yellow/purple surface
shown in Fig.~\ref{critical}; this surface represents observer
positions at which the source momentarily approaches
with an instantaneous speed of $c$ and with zero acceleration.
The surface can be found (with some effort) by differentiating Eq.~\ref{RtyFarty}
and setting $({\rm d}R/{\rm d}t)=-c$ and $({\rm d}^2R/{\rm d}t^2) = 0$.
With increasingly large distances [$(r_{\rm P}^2+z_{\rm P}^2)^{1/2} \gg r$],
the surfaces asymptotically tend to cones with half angles
$\theta_{\rm P}= \sin^{-1}(c/v)$, where
$v=r\eta$ is the instantaneous (tangential) speed of the source.
The green and dark blue curves in Fig.~\ref{flatbits}
correspond to such an observation angle.

Inside the surface (Fig.~\ref{critical}), plots of Eq.~\ref{C:3} will be similar to the red curve 
in Fig.~\ref{flatbits}, whilst outside it, an oscillatory form exemplified by the light blue curve
occurs.

The discussion in the final section of the main paper focuses on $t_{\rm P}~{\rm versus}~t$ curves
such as the green and dark blue examples shown in Fig.~\ref{flatbits}.
Apart from a single point at their center where ${\rm d}t_{\rm P}/{\rm d}t=0$,
the ``steps'' are not flat~\cite{dissert}. However, there is a reasonable region
of $t$ over which ${\rm d}t_{\rm P}/{\rm d}t \ll 1$.


\begin{thebibliography}{99}
\bibitem{footnotexz}
As mentioned above, the $x-$ and $z-$ extent of the antenna are small compared to its length. 
Therefore we ignore the slight variations in distance caused by the non-zero $x$ depth and 
$z$ height of the antenna and represent the motion of the volume-distributed polarization 
current by a function depending only on $y$ and $t$.
\end{thebibliography}
\end{document}